\UseRawInputEncoding
\documentclass[twocolumn,amsmath,amssymb,nofootinbib,longbibliography]{revtex4-2}

\usepackage{graphicx}
\usepackage{dcolumn}
\usepackage{bm}
\usepackage{color}
\usepackage{lipsum}
\usepackage{ulem}
\usepackage{hhline}
\usepackage[a4paper]{hyperref}
\usepackage{mwe}
\hypersetup{colorlinks=true,linktoc=all,linkcolor=blue,breaklinks=true,citecolor=blue,urlcolor=blue}

\newcommand{\be}{\begin{equation}}
\newcommand{\ee}{\end{equation}}
\newcommand{\bea}{\begin{eqnarray}}
\newcommand{\eea}{\end{eqnarray}}

\usepackage{natbib}

\begin{document}

\title{Spin waves in bilayers of transition-metal dichalcogenides}

\author{Wojciech Rudzi\'nski$^{1}$}
\email{wojrudz@amu.edu.pl}

\author{J\'ozef Barna\'s$^{1,2}$}
\author{Anna Dyrda\l$^{1}$}

\affiliation{$^{1}$Faculty of Physics, Adam Mickiewicz University in Pozna\'n, ul. Uniwersytetu Pozna\'nskiego 2, 61-614 Pozna\'n, Poland},

\affiliation{$^{2}$Institute of Molecular Physics, Polish Academy of Sciences,\\ ul. M. Smoluchowskiego 17,	60-179 Pozna\'n, Poland}

\date{\today}
\begin{abstract}
Van der Waals magnetic materials are currently of great interest as materials for applications in future ultrathin nanoelectronics and nanospintronics. Due to weak coupling between individual monolayers, these materials can be easily obtained in the monolayer and bilayer forms. The latter are of specific interest as they may be considered as  natural two-dimensional spin valves. In this paper,  we study theoretically spin waves in bilayers of transition metal dichalcogenides. The considerations are carried within the  general spin wave theory based on effective spin Hamiltonian and Hollstein-Primakoff-Bogolubov transformation. The spin Hamiltonian includes intra-layer as well as inter-layer  nearest-neighbour exchange interactions, easy-plane anisotropy,  and additionally a weak in-plane easy-axis anisotropy. The bilayer systems consist of two ferromagnetic (in-plane magnetization) monolayers that are coupled either ferromagnetically or antiferromagnetically. In the latter case, we analyse the spin wave spectra in all magnetic phases, i.e. in the antiferromagnetic, spin-flop, and ferromagnetic ones.

\end{abstract}
\maketitle

\section{Introduction}
Two-dimensional (2D) van-der-Waals magnetic materials are currently of great interest due to expected applications in atomically-thin spintronics devices, like for instance spin valves, memory elements, and others~\cite{Burch2018,Wang22,Li2022,jafariJMMM}. In the bulk form, these materials are build of monolayers that are weakly coupled by van-der-Waals forces. Therefore, it is relatively easy to obtain them in the form of thin films with arbitrary number of monolayers, down to bilayers and  single monolayers.  Accordingly,  magnetic ordering and magnetic properties of van der Waals materials depend on the  number of coupled monolayers.

Expected applications of 2D magnetic materials stimulated also intensive  theoretical investigations of their physical properties as well as search for new materials with better characteristics, especially with higher Curie/Neel temperatures. Of particular interest are their electronic and magnetic properties, including also spin dynamics and spin wave propagation \cite{23,per,jain,chen,costa,ortm,6}.

Magnetic ground state off van der Waals materials can be relatively easily tuned by external strain or gating~\cite{Lu2020,Cui2020,Verzhbitskiy2020,Tan2021}. These properties, together with strong magnetoresitance effects, magneto-optical properties, spin-filtering, spin-to-charge interconversion, and topological (electronic and magnon) transport make these materials  extremely attractive not only for applications but also for theoretical studies.
Moreover,  magnetic van-der-Waals structures  also offer unique possibilities for tuning magnetic anisotropy~\cite{7}, which is crucial for various applications where magnetic anisotropy plays  an active role. Magnetic anisotropy also allows to overcome large spin fluctuations in 2D systems and therefore facilitates stabilization of the spin structure. Interestingly, such a tuning of  magnetic anisotropy in van der Waals materials can be easily  achieved with chemical doping, externally-induced strains, or proximity effects \cite{Burch2018}.

Various groups of magnetic van der Waals materials are currently known. These materials have also various physical and especially transport properties, including ferromagnetic semiconductors, e.g. VS$\textsubscript{2}$, VSe$\textsubscript{2}$
\cite{8,9,10,11,12},
itinerant ferromagnets such as Fe$\textsubscript{3}$GeTe$\textsubscript{2}$ \cite{Wang22,brataas}, and insulating ferromagnets like CrI$\textsubscript{3}$ \cite{huang} or Cr$\textsubscript{2}$Ge$\textsubscript{2}$Te$\textsubscript{6}$ \cite{Burch2018,gong}.
In this paper we focus on magnetic properties and especially on
spin wave excitations in  a specific group of ferromagnetic 2D  van der Waals materials, i.e. in transition-metal dichalcogenides (TMDs) \cite{Chhowalla2013,Feng2011,Zhang2013,19}, MX$\textsubscript{2}$, where M stands for a transition metal atom and X  for a chalcogen one (X=S, Se, Te). These materials have Curie temperatures in the vicinity of room temperature, so they are of certain potential for practical applications.
We also note that spin wave excitations in van der Waals structures  are currently of great interest for magnonics applications,  and have been  studied theoretically as well as experimentally in various materials, including TMDs, chromium trihalides (CrI$_3$, CrCl$_3$)  and others. It has been shown, for instance, that spin waves in chromium trihalides have features that follow from  topological properties of these materials~\cite{chen,costa}. Such topologically-induced features also occur in TMDs \cite{16,17,18}.

A specific subgroup of TMDs are Vanadium-based dichalcogenides, VX$_2$ with X=S, Se and Te~\cite{8,9,10,11,12,Chhowalla2013,Feng2011,Zhang2013}. Two different polymorphs of  VX$_{2}$ materials are currently known, i.e.  the trigonal prismatic crystallographic structure (shortly referred to in the following as the H phase) and the octahedral structure (referred to as the T phase)~\cite{Chhowalla2013,Feng2011,Zhang2013}.
An individual monolayer of these materials  consists of a hexagonal atomic plane of Vanadium atoms, sandwiched between two chalcogen (X) atomic planes. In-plane positions of the chalcogen atoms in these two planes are the same (one on the other) in the T phase but these planes are rotated by some angle in the H phase.
Experimental and theoretical studies reveal a metal-insulator transition~\cite{9} when reducing thickness of  VSe$\textsubscript{2}$ layers down to a 2D monolayer.
It turn, the intrinsic ferromagnetism in the VSe$\textsubscript{2}$ monolayers has been reported in a number of experimental as well as theoretical works
(see e.g. \cite{9,10,11}).

\begin{figure*}[hbt!]
\centering
\includegraphics[width=2.2\columnwidth]{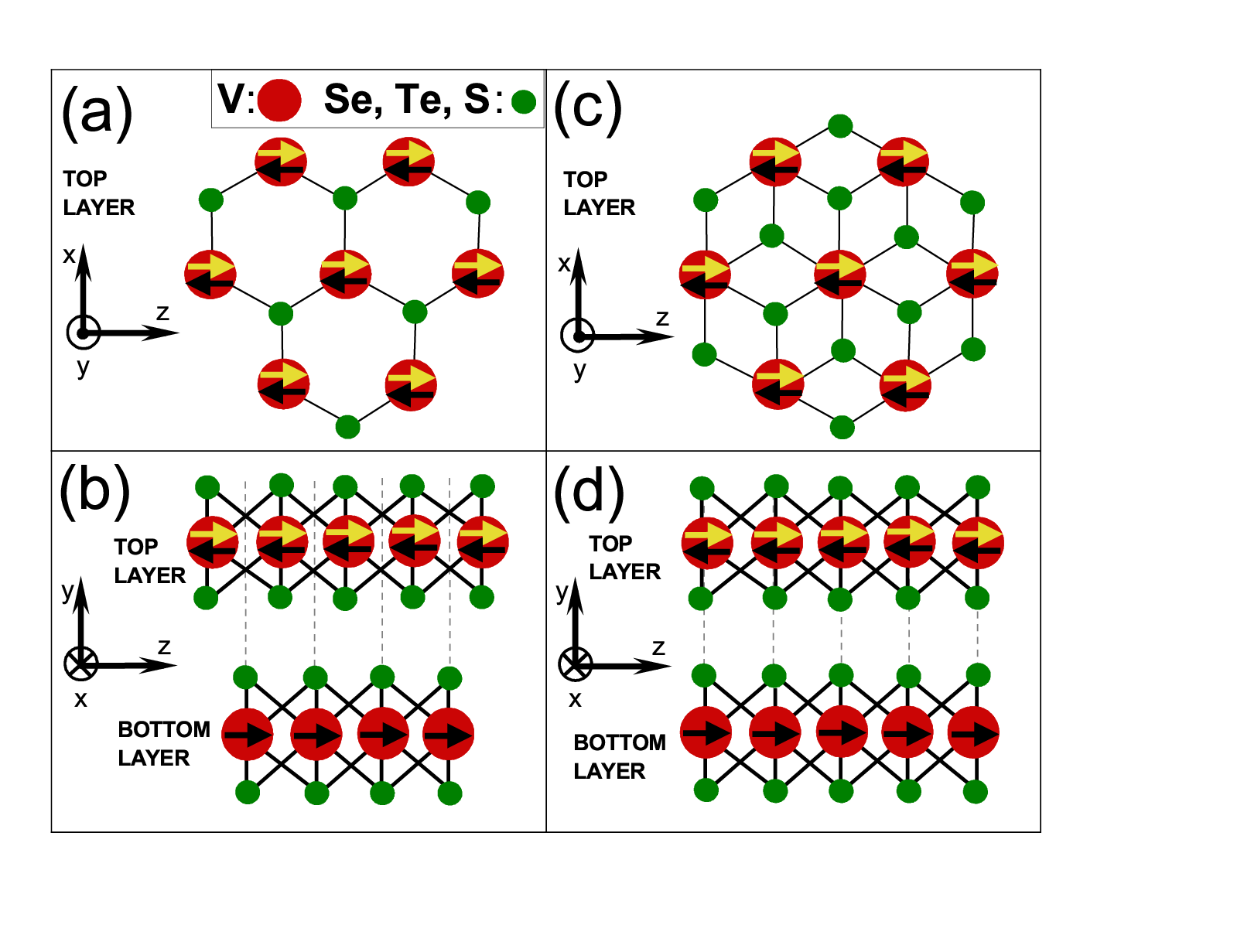}\hfil
\caption{Top view  of the VX$\textsubscript{2}$ (X= Se, Te, S) monolayer in the H-phase (a) and T-phase (c). The monolayer is in the ($x,z$) plane while the axis $y$ is normal to the plane. The side view of the corresponding bilayer systems along the z-axis is shown in (b) and (c) for the H and T phases, respectively. Ground state spin orientation of the bottom layer is assumed  along the $z$-axis while that of the top layer is along $-z$ axis for the antiferromagnetically coupled bilayers (black arrows) and along $z$ axis for the ferromagnetic bilayers (yellow arrows).
For the trigonal prismatic  structure (H-phase), one monolayer sits on the other with $\pi$ rotation, which implies that each vanadium atom (see the position of the red dots)  has three NNs  in the adjacent monlayer. For the octahedral  structure (T-phase) of VX$_2$, each vanadium atom  has one NN  in the adjacent monolayer.}
    \label{Fig:5}
\end{figure*}

Individual monolayers of TMDs are usually ferromagnetic with the magnetization oriented in the layer plane, though systems with magnetization normal to the plane can also occur. Moreover, antiferromagnetically arranged monolayers are possible as well.
Of particular interest seem to be the  bilayer structures, especially those with ferromagnetic monolayers coupled antiferromagnetically  due to  interlayer exchange coupling. This is because such bilayers in an external magnetic filed may  be considered as natural atomically thin spin valves~\cite{jafariJMMM}, and may be a fundamental building blocks of two dimensional spintronics.

In this paper we analyse  spin wave excitations  in the H-stacked as well as T-stacked bilyers of TMDs, with particular attention to the
interplay of the   magnetocrystalline anisotropy, intralayer ferromagnetic exchange coupling, and interlayer antiferromagnetic or ferromagnetic coupling. In order to find the magnon spectra we use the relevant spin Hamiltonian  and apply the Holstein-Primakoff transformation followed by the diagonalization procedure based on the  Bogolubov transformation~\cite{brataas,kamra,brandes,owerre}.

Vanadium-based dichalcogenides usually display an easy-plane single-ion magnetic anisotropy -- though easy-axis single-ion anisotropy normal to the plane has also been found owing to the proximity effects~\cite{Abdollahi2023}. Such an anisotropy has been proven by many DFT numerical calculations and also evaluated experimentally, so its presence in TMDs is now well established~\cite{10,11,Liu2014,Wang2021,Liu2023,Esters2017,Abdollahi2023} -̶  even though a single-ion  magnetic anisotropy for spin 1/2 model was believed to vanish.  Moreover, it was shown that this magnetic anisotropy can be tuned externally, eg. by a strain~\cite{Feng2018}.

In turn, the esistence of an in-plane easy-axis anisotropy is not so clear. Such an anisotropy was found in Ref.~\cite{10}, however it was about two orders of magnitude smaller than the corresponding easy-plane anisotropy.
On one hand,  existence of such an in-plane easy-axis single-ion anisotropy is believed to vanish for spin 1/2 models, and also on symmetry arguments. However, there are several ingredients that may effectively contribute to such an anisotropy. First of all, exact value of the magnetic moments of V atoms in V-based TMDs deviates from 1$\mu_B$, as follows from DFT calculations and also from experimental measurements, indicating that these materials cannot not be described by strictly spin 1/2 models. Secondly, some other contributions may be due to magnetic dipolar interactions, spin-orbit interaction associated with mobile electrons, proximity to a substrate, and external strains. Especially the latter seems to be very effective and important.

Therefore, in our paper we assume the easy-plane anisotropy, and following Ref.~\cite{10}, we also assume a small in-plane easy-axis anisotropy (two orders of magnitude smaller than the easy-plane anisotropy, as in Ref.~\cite{10}. Though the later anisotropy is rather difficult to be detected experimentally, especially in the spin-wave spectrum, we include it to have a general situation and general formulas that can be used if necessary for materials where such an anisotropy is larger, especially that this anisotropy is also  tunable externally. Therfore, one may expect that some data on higher values of the easy-axis anisotropy will be available in the near future, e.g. in strained systems. However, for numerical calculations of spinvave modes we  assume the value given in Ref.~\cite{10}, and the role  of easy-plane  and in-plane easy-axis anisotropies is analyzed in detail.  

The model and theoretical method used to study spin wave spectra are described in more details in section 2, where the spin Hamiltonian used to describe TMDs is defined. Numerical results are presented  and discussed in section 3. Summary and final conclusions are in section 4.

\section{Antiferromagnetically coupled bilayers}
\label{Sec:IIA}
We consider the case when the perpendicular anisotropy is of easy-plane type, and assume the coordinate system with the axis $y$ normal to the layers and the axis $z$ along the in-plane easy axis.  Though the in-plane easy-axis anisotropy is rather small in Vanadium dichalcogenides, we consider a general model when this anisotropy may be remarkable and may play some  role. We also assume an external magnetic field along the easy axis.

To describe spin waves in a bilayer consisting of two ferromagnetic monolayers coupled antiferromagetically we  use the following model spin Hamiltonian:
\begin{equation}
    H=\sum_\alpha H^{\alpha}+H_{\rm int},
\end{equation}
where $\alpha$=T ($\alpha$=B) stands the top (bottom) monolayer,
and Hamiltonian of the $\alpha$th monolayer includes three terms,
$H^{\alpha}=H^{\alpha}_{\rm ex}+H^{\alpha}_A+H^{\alpha}_h$. Here, the first term stands for
the ferromagnetic intralayer exchange coupling, the second term includes the magnetic anisotropies in the system, and the last term is the Zeemann energy in an external magnetic field $h$. The Hamiltonian $H^\alpha$ can be written explicitly in the form,
\begin{eqnarray}
    H^{\alpha}=J_1\sum_{\textbf{r},\bm{\delta}}\textbf{S}_{\textbf{r},\alpha}\cdot\textbf{S }_{\textbf{r}+\bm{\delta},\alpha} + \frac{D_y}{2}\sum_{\textbf{r}}\Big(S_{\textbf{r},\alpha}^y\Big)^2 \nonumber \\ -\frac{D_z}{2}\sum_{\textbf{r}}\Big(S_{\textbf{r},\alpha}^z\Big)^2 -h\sum_{\textbf{r}}S_{\textbf{r},\alpha}^z .
\end{eqnarray}
Here, the exchange coupling between Vanadium atoms within the monolayer is ferromagnetic, $J_1<0$, the easy-plane anisotropy constant $D_y$ and the  in-plane easy-axis anisotropy constant $D_z$ are both defined as positive, while the magnetic field $h$ is taken in energy units ($g\mu_B h\Rightarrow h$, where $g$ denotes the gyromagnetic factor and $\mu_B$ is the Bohr magneton).
The summation over $\textbf{r}$ denotes here the summation over lattice sites, while that over $\bm{\delta}$ is the summation over nearest neighbours, with $\bm{\delta}$ standing for the vectors connecting a particular site to its in-plane nearest neighbours (NNs). [We neglect here exchange coupling between next-nearest neighbours.]   In turn, the last term in Eq.(1) describes the antiferromagnetic exchange coupling between the two monolayers,
\begin{equation}
    H_{int}=2J_2\sum_{\textbf{r},\bm{\delta}}\textbf{S}_{\textbf{r},T}\cdot\textbf{S}_{\textbf{r}+\bm{\delta},B},
\end{equation}
with $J_2>0$.  Here, the summation is over lattice sites $\textbf{r}$ in one monolayer only (therefore, there is a factor of 2 in front on the right side). Note, $\bm{\delta}$ corresponds here to inter-layer NNs (NNs in the adjacent layer).

When the above defined system is in an external magnetic field applied along the in-plane easy axis ($z$ axis), one may expect in general three different stable spin configurations in the bilayer; (i) antiferromagnetic state with the spins of the bottom layer  oriented say along $+z$ axis and of the top layer along $-z$ axis, respectively, (ii) spin-flop (canted) phase with the spins of the two monolayers oriented in the atomic planes at an angle $\chi$ to the $z$ axis, and (iii)   the ferromagnetic  phase with the spins of both layers oriented along the  $z$ axis (this corresponds to $\chi =0$).
The transition from antiferromagnetic to spin-flop  phase occurs at $h=h_{\rm {sf}}$ (see Appendix A), with
\begin{equation}
    h_{\rm sf}=S\sqrt{D_z(4{\xi}J_2-D_z)}
\end{equation}
for $4{\xi}J_2-D_z >0$,
where  $\xi$ denotes the structure factor, $\xi =3$ and $\xi =1$ for the H and T phases, respectively.
The spin-flop phase appears in the range of magnetic fields $h_{\rm {sf}}< h < h_s$,
where $h_s$ is the threshold magnetic field (the saturation field), at which the transition from the  spin-flop  phase to the ferromagnetic  one  occurs (see Appendix A),
\begin{equation}
h_s=S(4{\xi}J_2-D_z).
\end{equation}
If $D_z=0$, then $h_{\rm sf}=0$ and $h_s=4S{\xi}J_2$.   Note, that in a general case the spin-flop phase appears when $4{\xi}J_2>D_z$, while in the opposite case, $4{\xi}J_2<D_z $, there is a direct (metamagnetic) transition) from the antiferromagnetic to ferromagnetic phase (there is then no spin-flop phase). In the systems considered here $D_z$ is very small, much smaller than $J_2$. Accordingly, the spin-flop field $h_{\rm {sf}}$  is also very small, so there is no metamagnetic transition and the antiparallel configuration may occur in a very narrow range of magnetic field and also at very low temperatures.

\subsection{Spin waves in the antiferromagnetic phase}
\label{Sec:IIC}

We consider first the antiferromagnetic (AF) phase, which appears below the transition field to the spin-flop phase. The spin moments of the bottom layer are along $z$ axis, while of the top layer are along the $-z$ axis, see Fig.1.  In the first step we perform the Holstein-Primakoff transformation:
\begin{eqnarray}
    S_{\textbf{r},T}^x=\sqrt{\frac{S}{2}}(a_{\textbf{r},T}^++a_{\textbf{r},T}),\nonumber\\
    S_{\textbf{r},T}^y=i\sqrt{\frac{S}{2}}(a_{\textbf{r},T}-a_{\textbf{r},T}^+),\nonumber\\
    S_{\textbf{r},T}^z=a_{\textbf{r},T}^+a_{\textbf{r},T}-S,
\end{eqnarray}
for  spins oriented antiparallel to the z-axis (top layer), and
\begin{eqnarray}
    S_{\textbf{r},B}^x=\sqrt{\frac{S}{2}}(a_{\textbf{r},B}^++a_{\textbf{r},B}),\nonumber\\
    S_{\textbf{r},B}^y=i\sqrt{\frac{S}{2}}(a_{\textbf{r},B}^+-a_{\textbf{r},B}),\nonumber\\
    S_{\textbf{r},B}^z=S-a_{\textbf{r},B}^+a_{\textbf{r},B},
\end{eqnarray}
for spins oriented along the z-axis (bottom layer). Here,  $a_{\textbf{r},\alpha}^+$  ($a_{\textbf{r},\alpha}$) is the bosonic  creation (anihilation) operator.

Upon inserting the Holstein-Primakoff transformation into Eqs. (1) to (3), keeping terms up to the second order in the magnon operators and disregarding any constant terms, one finds,
\begin{eqnarray}
    H & = & J_1S\sum_{\textbf{r},\bm{\delta},\alpha}\Big(a_{\textbf{r},\alpha}^+a_{\textbf{r},\alpha}+a_{\textbf{r}+\bm{\delta},\alpha}^+a_{\textbf{r}+\bm{\delta},\alpha}
    \nonumber\\
    & & -a_{\textbf{r},\alpha}^+a_{\textbf{r}+\bm{\delta},\alpha}-a_{\textbf{r}+\bm{\delta},\alpha}^+a_{\textbf{r},\alpha}\Big)
    +2J_2S\sum_{\textbf{r}}\Big[\sum_{\alpha}a_{\textbf{r},\alpha}^+a_{\textbf{r},\alpha}
    \nonumber\\
    & &
    +\sum_{\bm{\delta}}\Big(a_{\textbf{r},T}^+a_{\textbf{r}+\bm{\delta},B}^++a_{\textbf{r},T}a_{\textbf{r}+\bm{\delta},B}\Big)\Big]
    \nonumber\\
    & & -\frac{D_yS}{4}\sum_{\textbf{r},\alpha}\Big(a_{\textbf{r},\alpha}^+a_{\textbf{r},\alpha}^++a_{\textbf{r},\alpha}a_{\textbf{r},\alpha}-2a_{\textbf{r},\alpha}^+a_{\textbf{r},\alpha}\Big)
    \nonumber\\
    & &
    +D_zS\sum_{\textbf{r},\alpha}a_{\textbf{r},\alpha}^+a_{\textbf{r},\alpha}
    -h\sum_{\textbf{r}}\Big(a_{\textbf{r},B}^+a_{\textbf{r},B}-a_{\textbf{r},T}^+a_{\textbf{r},T}\Big).
    \nonumber\\
    & &
\end{eqnarray}
Then we perform the Fourier transformation to the momentum space,
\begin{eqnarray}
    &&a_{\textbf{r},\alpha}=\frac{1}{\sqrt{N}}\sum_{\textbf{k}}a_{\textbf{k},\alpha}e^{-i\textbf{k}{\cdot}\textbf{r}},
    \nonumber\\
    & &
    a_{\textbf{r},\alpha}^+=\frac{1}{\sqrt{N}}\sum_{\textbf{k}}a_{\textbf{k},\alpha}^+e^{i\textbf{k}{\cdot}\textbf{r}},
\end{eqnarray}
where N is the number of unit cells, and  $\textbf{k}$ is the wavevector from the first Brillouin zone, see Appendix B.
Upon this transformation
one may rewrite Hamiltonian, Eq. (8), in the following form:
\begin{eqnarray}
    H & = & \sum_{\textbf{k}}\bigg{\{}2J_1S\sum_{\alpha}(\gamma_\textbf{k}-6)a_{\textbf{k},\alpha}^+a_{\textbf{k},\alpha}
    \nonumber\\
    & & +2J_2S\Big(\sum_{\alpha}{\xi}a_{\textbf{k},\alpha}^+a_{\textbf{k},\alpha}+\eta_{\textbf{k}}a_{-\textbf{k},T}^+a_{\textbf{k},B}^++\eta_{\textbf{k}}^*a_{-\textbf{k},T}a_{\textbf{k},B}\Big)
    \nonumber\\
    & &+S\sum_{\alpha}\Big[-\frac{D_y}{4}(a_{\textbf{k},\alpha}^+a_{-\textbf{k},\alpha}^++a_{\textbf{k},\alpha}a_{-\textbf{k},\alpha}-2a_{\textbf{k},\alpha}^+a_{\textbf{k},\alpha})
    \nonumber\\
    & &
    +D_za_{\textbf{k},\alpha}^+a_{\textbf{k},\alpha}\Big]
    +h(a_{\textbf{k},B}^+a_{\textbf{k},B}-a_{\textbf{k},T}^+a_{\textbf{k},T})\bigg{\}},
\end{eqnarray}
where the quantity $\xi$  is defined below  Eq. (4), while the structure factors $\gamma_\textbf{k}$  and $\eta_\textbf{k}$ read
\begin{equation}
    \gamma_\textbf{k}=2\bigg(\cos (k_za)+2\cos (\frac{\sqrt{3}}{2}k_xa) \cos (\frac{1}{2}k_za)\bigg),
\end{equation}
\begin{equation}
    \eta_{\bm{k}} = \left\{ \begin{array}{ll}
1 & \textrm{(for T phase)}\\
e^{i\frac{k_xa}{\sqrt{3}}}+2e^{-i\frac{k_xa}{2\sqrt{3}}}\cos ({\frac{1}{2}k_za)} & \textrm{(for H phase).}
\end{array} \right.
\end{equation}

As one can easily note, Eq.(10) may be written as
\begin{equation}
    H=H_{\textbf{k}}+H_{\textbf{-k}},
\end{equation}
where
\begin{eqnarray}
    H_{\textbf{k}} & = & \sum_{\textbf{k}}\bigg[\bigg(\frac{A_{\textbf{k}}^+}{2}\bigg)a_{\textbf{k},B}^+a_{\textbf{k},B}+ \bigg(\frac{A_{\textbf{k}}^-}{2}\bigg)a_{\textbf{k},T}^+a_{\textbf{k},T}
    \nonumber\\
    & & +B_{\textbf{k}}a_{\textbf{-k},T}a_{\textbf{k},B}+ C\sum_{\alpha}a_{\textbf{k},\alpha}a_{\textbf{-k},\alpha}\bigg]+H.c.,
\end{eqnarray}
with the  coefficients $A_{\textbf{k}}^{\pm}$, $B_{\textbf{k}}$ and C  given by the formulae,
\begin{eqnarray}
    &&A_{\textbf{k}}^{\pm}=S\bigg[2J_1\big(\gamma_\textbf{k}-6\big)+2{\xi}J_2+\frac{D_y}{2}+D_z\bigg]\> {\pm}\> h,
    \nonumber\\
    & &
    B_{\textbf{k}}=2{\eta}^*_{\textbf{k}}J_2S,
    \nonumber\\
    & &
     C=-\frac{D_yS}{4}.
\end{eqnarray}

Following Refs~\cite{brataas,kamra,brandes,owerre}, we define now a new wave vector $\bm{\kappa}$ that runs over half of the vector space of $\textbf{k}$, and define the four-dimensional Bogolubov transformation to new bosonic operators $\Theta_{{\pm}\bm{\kappa},\mu}$ and $\Theta_{{\pm}\bm{\kappa},\mu}^+$, with $\mu=+,-$ indexing the two magnon modes (to be specified below). This transformation can be written as
\begin{equation}
    \bm{\Theta}_{\bm{\kappa}}=\bm{\hat{T_4}}\textbf{a}_{\bm{k}},
\end{equation}
where
\begin{equation}
\bm{\Theta}_{\bm{\kappa}}=
\left( \begin{array}{c}
\Theta_{\bm{\kappa},I\>\>\>\>\>}\\
\Theta_{\bm{\kappa},II\>\>\>} \\
\Theta_{\bm{-\kappa},I\>\>}^+  \\
\Theta_{\bm{-\kappa},II}^+ \\
\end{array} \right), \hspace{1cm}
\textbf{a}_{\textbf{k}}=
\left( \begin{array}{c}
a_{\bm{k},T\>\>\>}\\
a_{\bm{k},B\>\>\>} \\
a_{\bm{-k},T}^+  \\
a_{\bm{-k},B}^+ \\
\end{array} \right),
\end{equation}
and $\bm{\hat{T_4}}$  denotes a  $2\mathcal{N}\times2\mathcal{N}$ paraunitary matrix  ($\mathcal{N}$ is a number of internal degrees of freedom within the unit cell), which  obeys
\begin{equation}
    [\bm{\Theta}_{\bm{\kappa}},\bm{\Theta}_{\bm{\kappa}}^+]=\bm{\hat{T_4}}[\bm{a}_{\bm{k}},\bm{a}_{\bm{k}}^+]\bm{\hat{T_4}}^+=\bm{\hat{T_4}}\bm{\hat{\sigma_z}}\bm{\hat{T_4}}^+=\bm{\hat{\sigma_z}},
\end{equation}
with the diagonal matrix $(\bm{\hat{\sigma_z}})_{l,l'}=\delta_{l,l'}\sigma_l$, where $\sigma_l=1$ for $l\leq{\mathcal{N}}$ and $\sigma_l=-1$ otherwise. The requirement given by Eq. (18) follows from the bosonic commutation relations $[\Theta_{\bm{\kappa},\mu},\Theta_{\bm{\kappa}',{\mu}'}^+]=\delta_{\bm{\kappa},\bm{\kappa}'}\delta_{\mu,{\mu}'}$. As a consequence, Eq. (16) can be written in the form
\begin{equation}
    \bm{\Theta}_{\bm{\kappa}}=\sum_{\alpha}
    \left( \begin{array}{cc}
u_{I,\alpha{\>\>\>\>\>}} & v_{I,\alpha{\>\>\>\>\>}}\\
u_{II,\alpha{\>\>\>}} & v_{II,\alpha{\>\>\>}}\\
\tilde{u}_{I,\alpha{\>\>\>\>\>}} & \tilde{v}_{I,\alpha{\>\>\>\>\>}}\\
\tilde{u}_{II,\alpha{\>\>}} & \tilde{v}_{II,\alpha{\>\>}}\\
\end{array} \right)
\left( \begin{array}{c}
a_{\bm{k},{\alpha\>\>\>\>}}\\
a_{\bm{-k},{\alpha}}^+  \\
\end{array} \right),
\end{equation}
where the Bogolubov coefficients $u_{\mu,\alpha}$ and $v_{\mu,\alpha}$ are evaluated at $\bm{\kappa}$ while the coefficients $\tilde{u}_{\mu,\alpha}$ and $\tilde{v}_{\mu,\alpha}$ are evaluated at $\bm{-\kappa}$. Moreover, the relation (19) requires the  normalization
\begin{equation}
    \sum_{\alpha}\big(|u_{\mu,\alpha}|^2+|v_{\mu,\alpha}|^2\big)=1,\>\sum_{\alpha}\big(|\tilde{u}_{\mu,\alpha}|^2+|\tilde{v}_{\mu,\alpha}|^2\big)=1.
\end{equation}

This procedure finally diagonalizes the Hamiltonian,
\begin{equation}
    H=\sum_{\bm{\kappa,\mu}}{\Big(}\omega_{\bm{\kappa},\mu}\Theta_{\bm{\kappa},\mu}^+\Theta_{\bm{\kappa},\mu}+\omega_{\bm{-\kappa},\mu}\Theta_{\bm{-\kappa},\mu}^+\Theta_{\bm{-\kappa},\mu}\Big).
\end{equation}
Employing Eq. (21), one obtains the relation $[\Theta_{\bm{\kappa},\mu},H]=\omega_{\bm{\kappa},\mu}\Theta_{\bm{\kappa},\mu}$, from which Eqs. (14) and (19) lead to the eigenvalue problem
\begin{equation}
    \Lambda_{\bm{\kappa}}\bm{e}_\mu=\omega_{\bm{\kappa},\mu}\bm{e}_\mu,
\end{equation}
with
\begin{equation}
\Lambda_{\bm{\kappa}}=
    \left( \begin{array}{cccc}
A_{\bm{\kappa}}^- & 0 & -2C & -B_{\bm{\kappa}}^*  \\
0 & A_{\bm{\kappa}}^+ & -B_{\bm{\kappa}} & -2C  \\
2C & B_{\bm{\kappa}}^* & -A_{\bm{\kappa}}^- & 0\\
B_{\bm{\kappa}} & 2C & 0 & -A_{\bm{\kappa}}^+  \\
\end{array} \right),
\bm{e}_\mu=
\left( \begin{array}{c}
u_{\mu,T}\\
u_{\mu,B}\\
v_{\mu,T}\\
v_{\mu,B}\\
\end{array} \right).
\end{equation}
Note that $A_{\bm{\kappa}}^{\pm}=A_{\bm{-\kappa}}^{\pm}$, $B_{\bm{\kappa}}^*=B_{\bm{-\kappa}}$ and therefore we have $u_{\mu,\alpha}=\tilde{u}_{\mu,\alpha}$, $v_{\mu,\alpha}=\tilde{v}_{\mu,\alpha}$. Moreover, as $\omega_{\bm{\kappa},\mu}$ is a real quantity, we also have $\omega_{\bm{-\kappa},\mu}=\omega_{\bm{\kappa},\mu}$. Finally, from Eqs. (22) and (23) one finds the appropriate dispersion relation.

For convenience, in this dispersion relation we come back to the usual notation for the wavevector, and replace in this relation $\bm{\kappa}$  by $\bm{k}$, which makes no confusion and no ambiguity. Thus, we write the dispersion relation as
\begin{eqnarray}
    \omega_{\bm{k},\mu} & = & \bigg{\{}A_{\bm{k}}^2-|B_{\bm{k}}|^2-4C^2+h^2
    \nonumber\\
    & & {\pm} 2\bigg[4C^2|B_{\bm{k}}|^2+h^2\Big(A_{\bm{k}}^2-|B_{\bm{k}}|^2\Big)\bigg]^{\frac{1}{2}}\bigg{\}}^{\frac{1}{2}},
\end{eqnarray}
where ${A_{\bm{k}}\equiv}A_{\bm{k}}^{\pm}{\mp}h$, and the sign $+$ and $-$ in the fifth term on the right hand side of of Eq. (24) corresponds to the spin-wave mode index $\mu=+$ and $\mu=-$, respectively.
From this equation follows that
 nonzero easy-plane anisotropy, $D_y >0$, leads to splitting of the magnon spectrum in the absence of magnetic field,  $h=0$, into two branches, $\omega_{\bm{k},{\pm}}$.

In the center of the first Brillouin zone, $\bm{k}=\bm0$ (point $\Gamma$), one then finds
\begin{equation}
    \omega_{\bm{k}=0,{\pm}}= S\big[2{\xi}J_2\big(D_y{\pm}D_y+2D_z\big)+D_yD_z+D_z^2\big]^{\frac{1}{2}}.
\end{equation}
This formula clearly shows that splitting of the magnon modes at the $\Gamma$ point appears for a nonzero $D_y$.
In turn, a nonzero in-plane magnetic anisotropy, $D_z >0$, leads to an energy gap. If $D_y=0$ and $D_z>0$, then the energy gap at $\Gamma$ is given by
\begin{equation}
    \omega_{\bm{k}=0,{+}}=\omega_{\bm{k}=0,{-}}= S\big(4{\xi}J_2D_z+D_z^2\big)^{\frac{1}{2}}.
\end{equation}
This gap vanishes if $D_z=0$, i.e., a two-fold degenerated Goldstone mode, $\omega_{\bm{k}=0,{\pm}}=0$, then  appears.
If $D_z=0$ and $D_y>0$, then
\begin{equation}
    \omega_{\bm{k}=0,{+}}= S\big(4{\xi}J_2D_y\big)^{\frac{1}{2}},
\end{equation}
\begin{equation}
    \omega_{\bm{k}=0,{-}}= 0  \,\,\,\,\textrm{(Goldstone mode)}.
\end{equation}
\subsection{Spin waves in the spin-flop phase}
\label{Sec:IID}
In the spin-flop phase, $h_{\rm sf}<h<h_s$, one needs first to rotate the spin operators to the local quantization axes appropriate for the top (T) and bottom (B) layers  (see Appendix A). This transformation does not affect the ferromagnetic intralayer exchange term of the model Hamiltonian (1,2), while the  antiferromagnetic interlayer coupling term $H_{\rm int}$ as well as the anisotropy $H_A^\alpha$ and Zeemann $H_h^\alpha$ terms now become
\begin{eqnarray}
H_{\rm int}& = &
2J_2\sum_{\textbf{r},\bm{\delta}}[\cos2\chi(S_{\textbf{r},T}^zS_{\textbf{r}+\delta,B}^z-S_{\textbf{r},T}^xS_{\textbf{r}+\delta,B}^x)
\nonumber\\
    & &-S_{\textbf{r},T}^yS_{\textbf{r}+\delta,B}^y],
\end{eqnarray}
\begin{eqnarray}
H_A^\alpha& = &
   \frac{1}{2} \sum_{\textbf{r}}\Big{\{}D_y(S_{\textbf{r},\alpha}^y)^2-D_z\big[(S_{\textbf{r},\alpha}^x)^2\sin^2\chi
    \nonumber\\
    & &+(S_{\textbf{r},\alpha}^z)^2\cos^2\chi\big]\Big{\}} ,
\end{eqnarray}
\begin{equation}
    H_{h}^{\alpha}=-h\cos\chi\sum_{\textbf{r}}S_{\textbf{r},\alpha}^z,
\end{equation}
where the spin operators are in the corresponding local systems.

Upon Holstein-Primakoff and Fourier transformations  (see Appendix C), one arrives at the Hamiltonian,  which -- bearing in mind the  Bogolubov transformation -- can be written as  $H=H_{\textbf{k}}+H_{\textbf{-k}}$, where
\begin{eqnarray}
    H_{\textbf{k}} & = & \sum_{\textbf{k}}\bigg[\sum_{\bm{\alpha}}\bigg(\frac{A_\textbf{k}}{2}\bigg)a_{\textbf{k},\alpha}^+a_{\textbf{k},\alpha}+B_{\textbf{k}}a_{\textbf{-k},T}a_{\textbf{k},B}
    \nonumber\\
    & & +\tilde{B}_{\textbf{k}}a_{\textbf{k},T}^+a_{\textbf{k},B}+ C\sum_{\alpha}a_{\textbf{k},\alpha}a_{\textbf{-k},\alpha}\bigg]+H.c.,
\end{eqnarray}
with
\begin{eqnarray}
    &&A_{\textbf{k}}=
    S\bigg[2J_1\big(\gamma_\textbf{k}-6\big)-2{\xi}J_2\cos2\chi+\frac{D_y}{2}
    \nonumber \\
    &&
    \hspace{0.9cm}+\frac{D_z}{2}\big(3\cos^2\chi-1\big)\bigg] +h\cos\chi ,
    \nonumber\\
   & &
    B_{\textbf{k}}=2{\eta}^*_{\textbf{k}}J_2S\sin^2{\chi},
\nonumber\\
    & &
    \tilde{B}_{\textbf{k}}=-2{\eta}^*_{\textbf{k}}J_2S\cos^2{\chi},
    \nonumber\\
    & &
     C=-\frac{S}{4}\big(D_y+D_z\sin^2{\chi}\big).
\end{eqnarray}
Then, the diagonalization of the Eq. (32), similarly as described in the preceding subsection, leads to the  eigenvalue problem, $\Lambda_{\bm{\kappa}}\bm{e}_\mu=\omega_{\bm{\kappa},\mu}\bm{e}_\mu$, where
\begin{equation}
\Lambda_{\bm{\kappa}}=
    \left( \begin{array}{cccc}
A_{\bm{\kappa}} & \tilde{B}^*_{\bm{\kappa}} & -2C & -B_{\bm{\kappa}}^*  \\
\tilde{B}_{\bm{\kappa}} & A_{\bm{\kappa}} & -B_{\bm{\kappa}} & -2C  \\
2C & B_{\bm{\kappa}}^* & -A_{\bm{\kappa}} & -\tilde{B}^*_{\bm{\kappa}}\\
B_{\bm{\kappa}} & 2C & -\tilde{B}_{\bm{\kappa}} & -A_{\bm{\kappa}}  \\
\end{array} \right),
\bm{e}_\mu=
\left( \begin{array}{c}
u_{\mu,T}\\
u_{\mu,B}\\
v_{\mu,T}\\
v_{\mu,B}\\
\end{array} \right).
\end{equation}

Upon replacing the notation  $\bm{\kappa}$  by $\bm{k}$,  we write the dispersion relation in  the form
\begin{eqnarray}
    \omega_{\bm{k},\mu} & = & \Big{\{}A_{\bm{k}}^2-|B_{\bm{k}}|^2+|\tilde{B}_{\bm{k}}|^2-4C^2
    \nonumber\\
    & & {\pm} \Big[-8A_{\bm{k}}C(B_{\bm{k}}^*\tilde{B}_{\bm{k}}+B_{\bm{k}}\tilde{B}^*_{\bm{k}}) +  B^2_{\bm{k}}\tilde{B}^{*2}_{\bm{k}}
     \nonumber\\
     & &+B^{*2}_{\bm{k}}\tilde{B}^2_{\bm{k}}+16|B_{\bm{k}}|^2C^2+4A_{\bm{k}}^2|\tilde{B}_{\bm{k}}|^2
     \nonumber\\
    & &-2|B_{\bm{k}}|^2|\tilde{B}_{\bm{k}}|^2  \Big]^{\frac{1}{2}}\Big{\}}^{\frac{1}{2}},
\end{eqnarray}
with the $+$ and $-$ signs in the fifth term on the right hand side of Eq. (35) corresponding to the mode index $\mu=+$ and $\mu=-$, respectively.
It is worth noting here that in case of the T-stacked geometry, both $B_{\bm{k}}$ and $\tilde{B}_{\bm{k}}$ are independent of $\bm{k}$,
$B_{\bm{k}}=B$ , $\tilde{B}_{\bm{k}}=\tilde{B}$, so that Eq. (35) simplifies to the following one:
\begin{equation}
    \omega_{\bm{k},\mu}  =  \Big{\{}(A_{\bm{k}}-\tilde{B})^2-(B\pm2C)^2\Big{\}}^{\frac{1}{2}}.
\end{equation}

At the saturation field $h_s$ and for $\bm{k}=\bm0$, the magnon energies  are equal
\begin{equation}
    \omega_{\bm{k}=0,{\mu}}= S\Bigg[\bigg(2{\xi}J_2{\pm}2{\xi}J_2+\frac{D_y}{2}\bigg)^2 -\bigg(\frac{D_y}{2}\bigg)^2\Bigg]^{\frac{1}{2}}.
\end{equation}
Thus, one finds
\begin{equation}
    \omega_{\bm{k}=0,{+}}= 2S\bigg[\big(2{\xi}J_2\big)^2+2{\xi}J_2\frac{D_y}{2} \bigg]^{\frac{1}{2}},
\end{equation}
\begin{equation}
    \omega_{\bm{k}=0,{-}}= 0 \,\,\,\, \textrm{(Goldstone mode)}.
\end{equation}
Thus, at the saturation field $h_s$, one of the modes is the Goldstone mode -- even in the case of nonzero in-plane easy-axis anisotropy, while the splitting of the modes (gap between the two modes) appears also in the case of vanishing $D_y$.  This behaviour is different from that in the antiferromagnetic state.

\subsection{Ferromagnetic state above the saturation field}

For $h\ge h_s$, the system is in the saturated (ferromagnetic) state, which can be considered as a special case of the spin-flop phase, corresponding to $\chi=0$. Equations (29) to (31) take then the forms
\begin{eqnarray}
H_{\rm int}& = &
2J_2\sum_{\textbf{r},\bm{\delta}}(S_{\textbf{r},T}^zS_{\textbf{r}+\delta,B}^z-S_{\textbf{r},T}^xS_{\textbf{r}+\delta,B}^x
\nonumber\\
    & &-S_{\textbf{r},T}^yS_{\textbf{r}+\delta,B}^y),
\end{eqnarray}
\begin{eqnarray}
H_A^\alpha& = &
   \frac{1}{2} \sum_{\textbf{r}}\Big[D_y(S_{\textbf{r},\alpha}^y)^2-D_z(S_{\textbf{r},\alpha}^z)^2\Big],
\end{eqnarray}
\begin{equation}
    H_{h}^{\alpha}=-h\sum_{\textbf{r}}S_{\textbf{r},\alpha}^z.
\end{equation}
Accordingly, the system Hamiltonian upon  Holstein-Primakoff and Fourier transformations acquires the form (32) with
 $A_{\textbf{k}}$, $B_{\textbf{k}}$, $\tilde{B}_{\textbf{k}}$   and $C$ given by the formulas
 \begin{eqnarray}
    &&A_{\textbf{k}}=
    S\bigg[2J_1\big(\gamma_\textbf{k}-6\big)-2{\xi}J_2+\frac{D_y}{2}+D_z\bigg]+h,
\nonumber\\
    & &
    B_{\textbf{k}}=0,
\nonumber\\
    & &
    \tilde{B}_{\textbf{k}}=-2{\eta}^*_{\textbf{k}}J_2S,
    \nonumber\\
    & &
     C=-\frac{SD_y}{4}.
\end{eqnarray}
Thus, the eigenvalue problem, $\Lambda_{\bm{\kappa}}\bm{e}_\mu=\omega_{\bm{\kappa},\mu}\bm{e}_\mu$, with
\begin{equation}
\Lambda_{\bm{\kappa}}=
    \left( \begin{array}{cccc}
A_{\bm{\kappa}} & \tilde{B}^*_{\bm{\kappa}} & -2C & 0  \\
\tilde{B}_{\bm{\kappa}} & A_{\bm{\kappa}} & 0 & -2C  \\
2C & 0 & -A_{\bm{\kappa}} & -\tilde{B}^*_{\bm{\kappa}}\\
0 & 2C & -\tilde{B}_{\bm{\kappa}} & -A_{\bm{\kappa}}  \\
\end{array} \right),
\bm{e}_\mu=
\left( \begin{array}{c}
u_{\mu,T}\\
u_{\mu,B}\\
v_{\mu,T}\\
v_{\mu,B}\\
\end{array} \right),
\end{equation}
leads to the final dispersion relation for the ferromagnetic  phase in the form
\begin{eqnarray}
    \omega_{\bm{k},\mu} =
    \Big[\big(A_{\bm{k}}\pm|\tilde{B}_{\bm{k}}|\big)^2-4C^2\Big]^{\frac{1}{2}}.
\end{eqnarray}

\section{Numerical results and discussion}
\label{Sec:III}
To discuss properties  of spin-wave spectra  in
2H-VX$_2$ bilayer systems,
we need first to fix the appropriate parameters, including the exchange and magnetic anisotropy constants, as well as the appropriate structural parameters.  As an example we will consider the spinwave spectra of a VTe$_2$ bilayer, and take the relevant parameters
from  DFT calculations~\cite{jafari}. Accordingly we assume: lattice parameter  $a=3.59 \AA$,  $D_y =3.812$ meV, $J_1=-14.85$ meV, and $J_2=0.19$  meV.
Since the model description is based on spin-1/2 Hamiltonian ($S=1/2$), the DFT-calculated  exchange parameters have been adapted accordingly. Only the  in-plane easy-axis anisotropy constant, $D_z=0.016$ meV, was taken from Ref.\cite{10}.
According to Eqs. (4) and (5), the corresponding
threshold magnetic fields $h_{\rm sf}$ and $h_s$ (converted from energy units to  Tesla) are $h_{\rm sf} =0.82$ T and $h_{\rm s} =9.65$ T.

\begin{figure}[hbt!]
\centering
    \includegraphics[width=1.0\columnwidth]{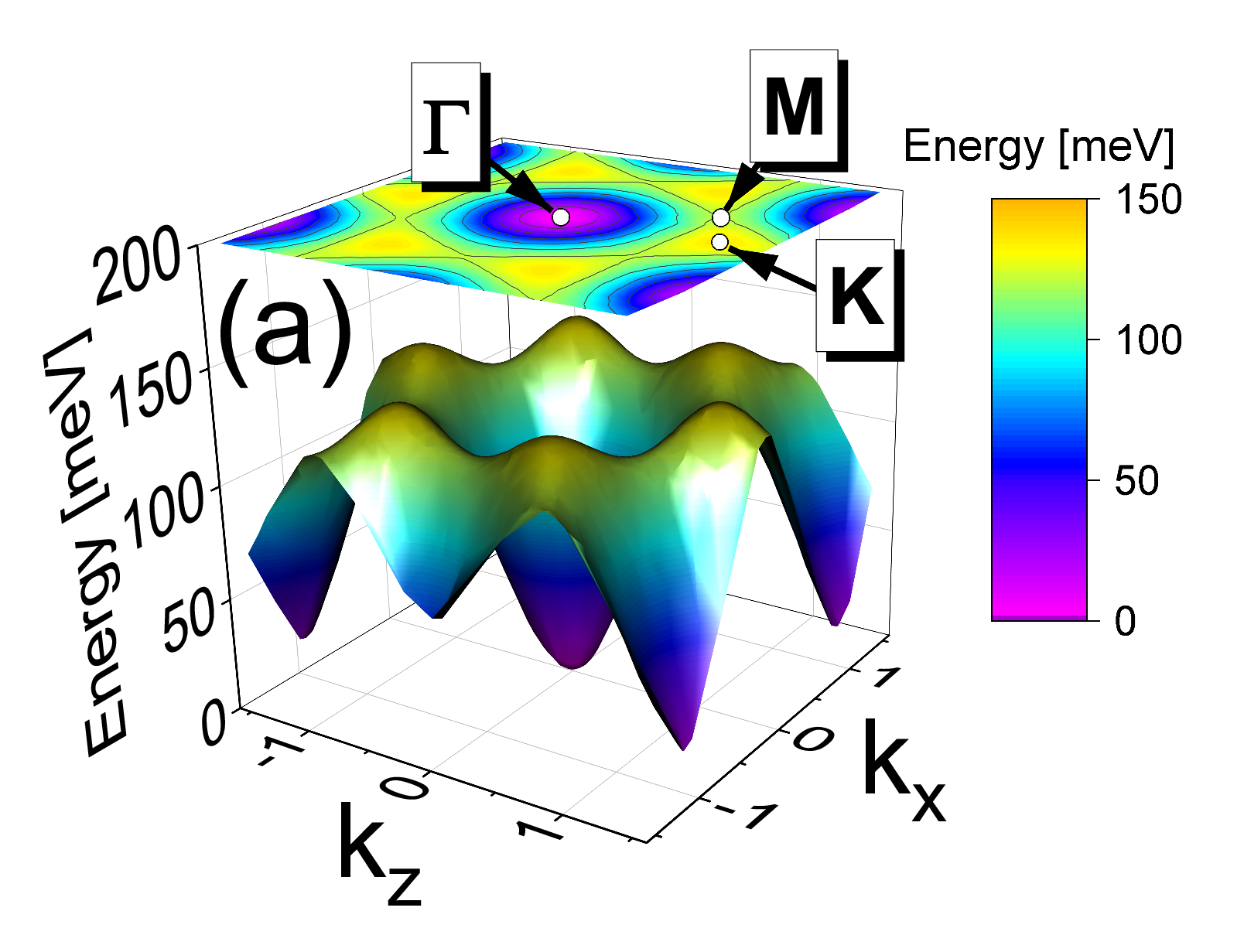}\\
    \includegraphics[width=1.0\columnwidth]{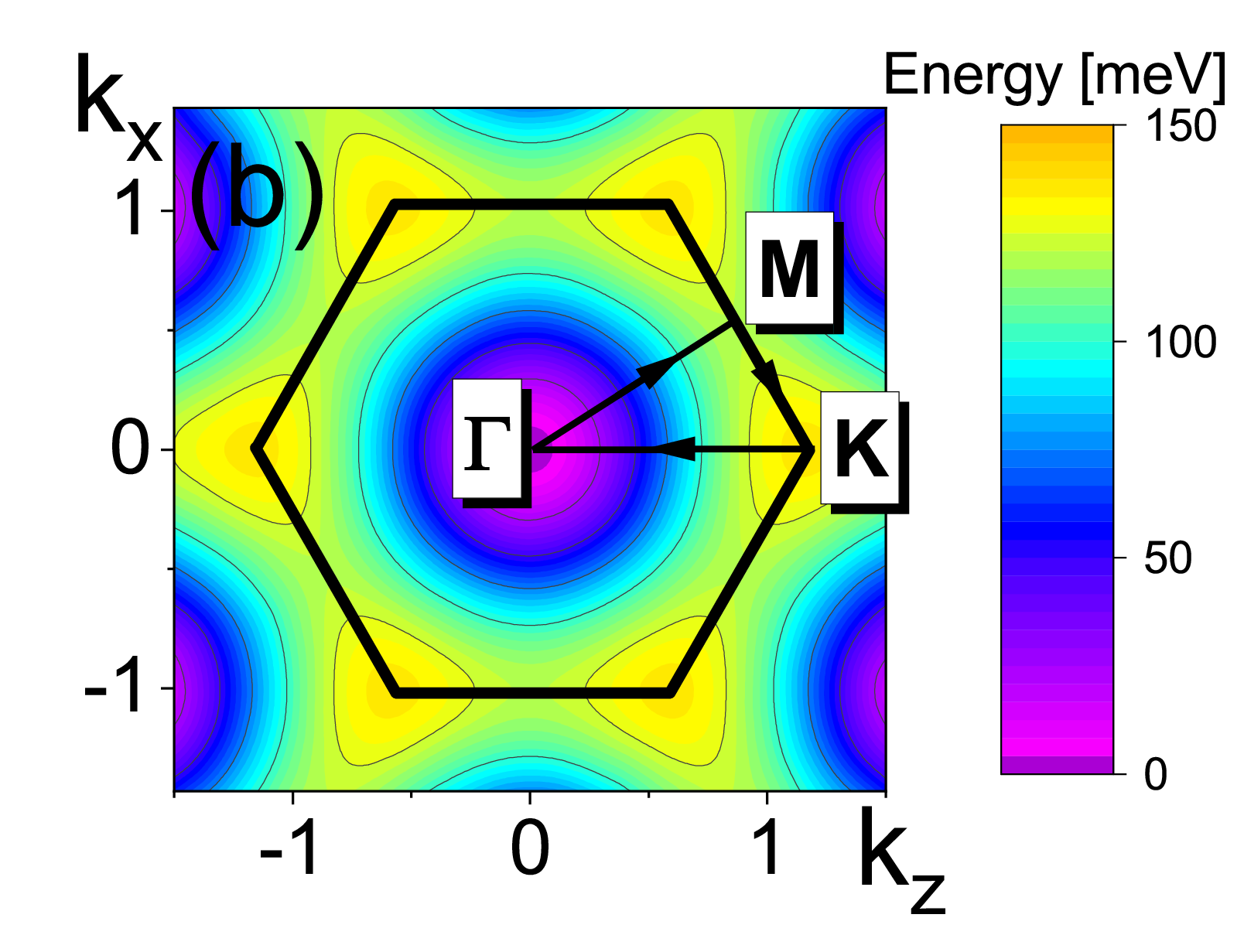}
\caption{The 3D view of the  magnon band in the first Brillouin zone, in the absence magnetic field ($h=$0), calculated for the  2H-VTe$_2$ bilayer (a) and its projection  onto the $(k_z,k_x)$-plane (b) with indicated Brillouin zone center ($\Gamma$) and high-symmetry points (K,M) and paths. See text for more details.}
    \label{Fig:2}
\end{figure}
\begin{figure}[hbt!]
\centering
    \includegraphics[width=0.95\columnwidth]{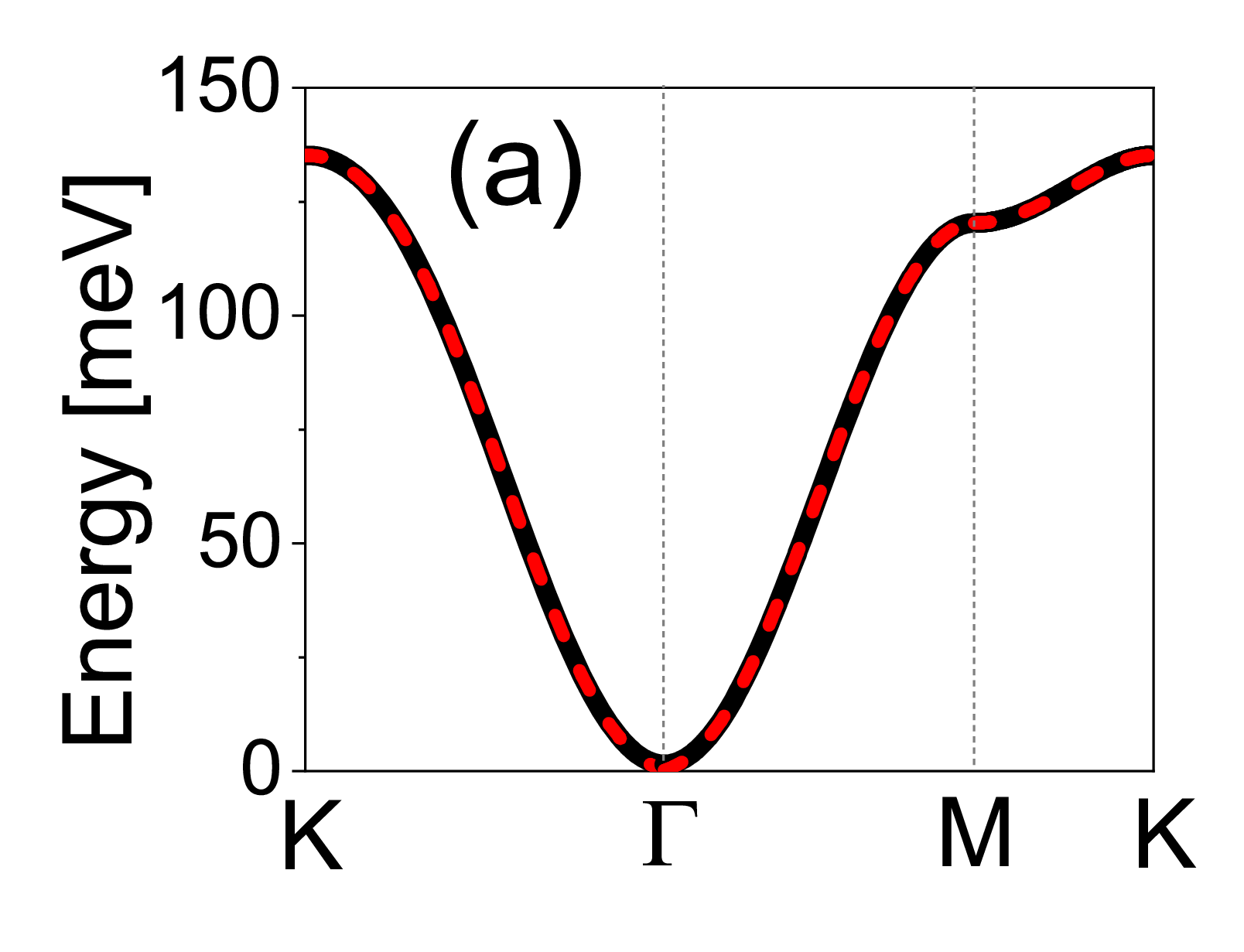}\\
    \includegraphics[width=0.99\columnwidth]{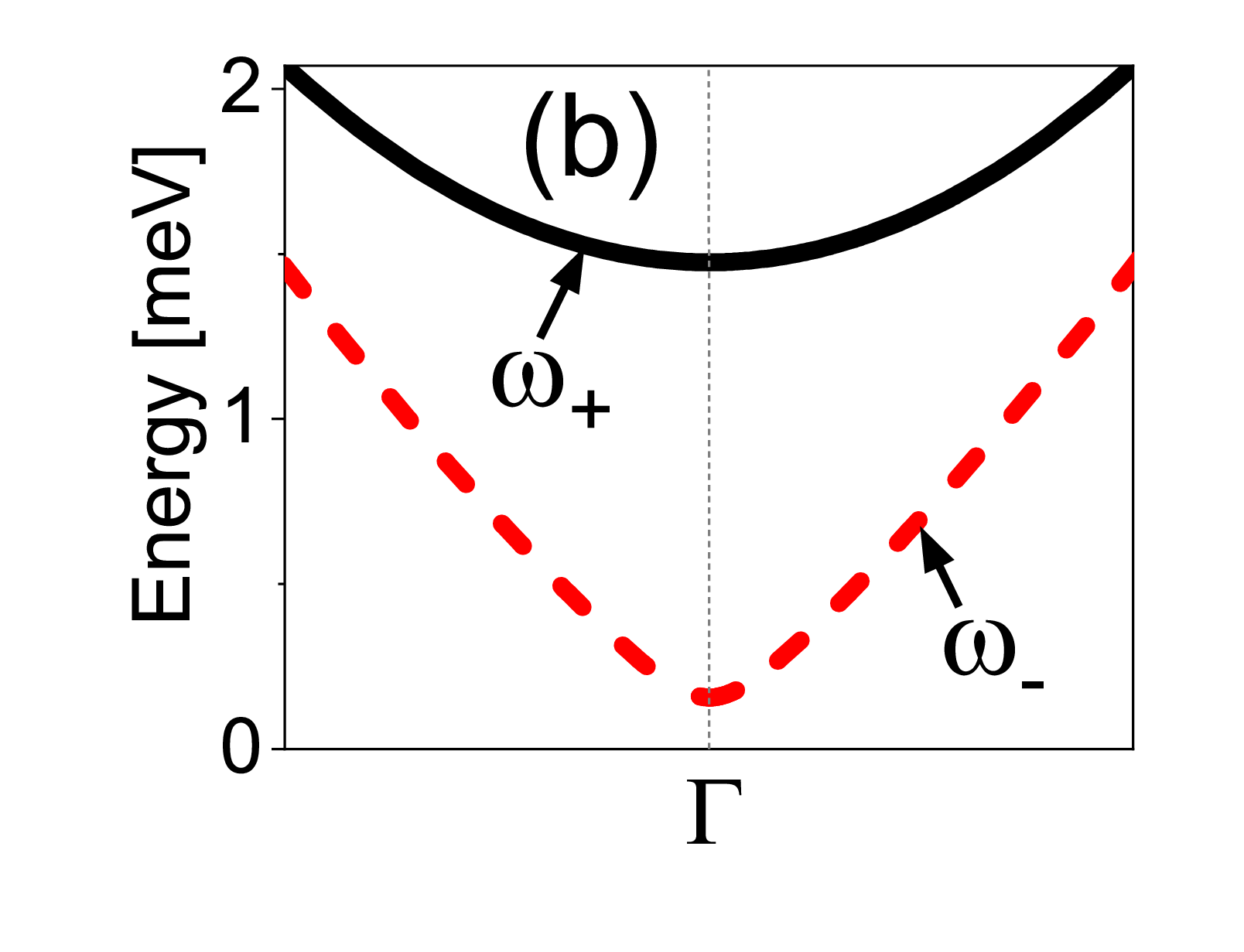}
\caption{Dispersion curves of the spin wave spectrum along the path K-$\Gamma$-M-K (a). As in Fig.2, the two modes are not resolved here. To observe splitting of the modes in we show in (b) the spin wave spectrum in a close vicinity of the $\Gamma$  point. Now, the splitting and also gap in the spectrum are clearly visible.}
    \label{Fig:3}
\end{figure}

The three-dimensional (3D) presentation of the magnon spectrum, calculated for the 2H-VTe$_2$ bilayer in the antiferromagnetic  phase and in the absence of external magnetic field (antiparalle configuration)   is shown in Fig.2a.  As there are two magnetic atoms in the  unit cell of the considered bilayer, one can expect two magnon bands. In the absence of magnetic field,  $h=0$, these two magnon modes have similar energy so they are not resolved in Fig.2a. However, even for $h=0$, these two modes can differ slightly in energy due to other  interactions, like interlayer exchange coupling and the easy-plane and in-plane easy-axis magnetic anisotropies. These interactions  introduce subtle effects (not resolved within the energy scale of Fig. 2a), which  may lead to a splitting of the spectrum into two spin-wave modes at zero external magnetic field, and also can generate a gap in the spectrum, ss will be shown below.
Low energy spin waves exist in the Brillouin zone center  around $\bm{k}=\bm 0$ (see the $\Gamma$ point in Fig. 2a), while the maxima of magnon energy appear at the
points K  of the Brillouin zone (see Fig. 2a,b).
The spin-wave energy for selected cross sections of the 3D magnon bands are displayed Fig.2b and Fig.3a along high-symmetry paths,  $K{\rightarrow}{\Gamma}{\rightarrow}M{\rightarrow}K$, in the momentum space.
\begin{figure}[hbt!]
\centering
    \includegraphics[width=0.95\columnwidth]{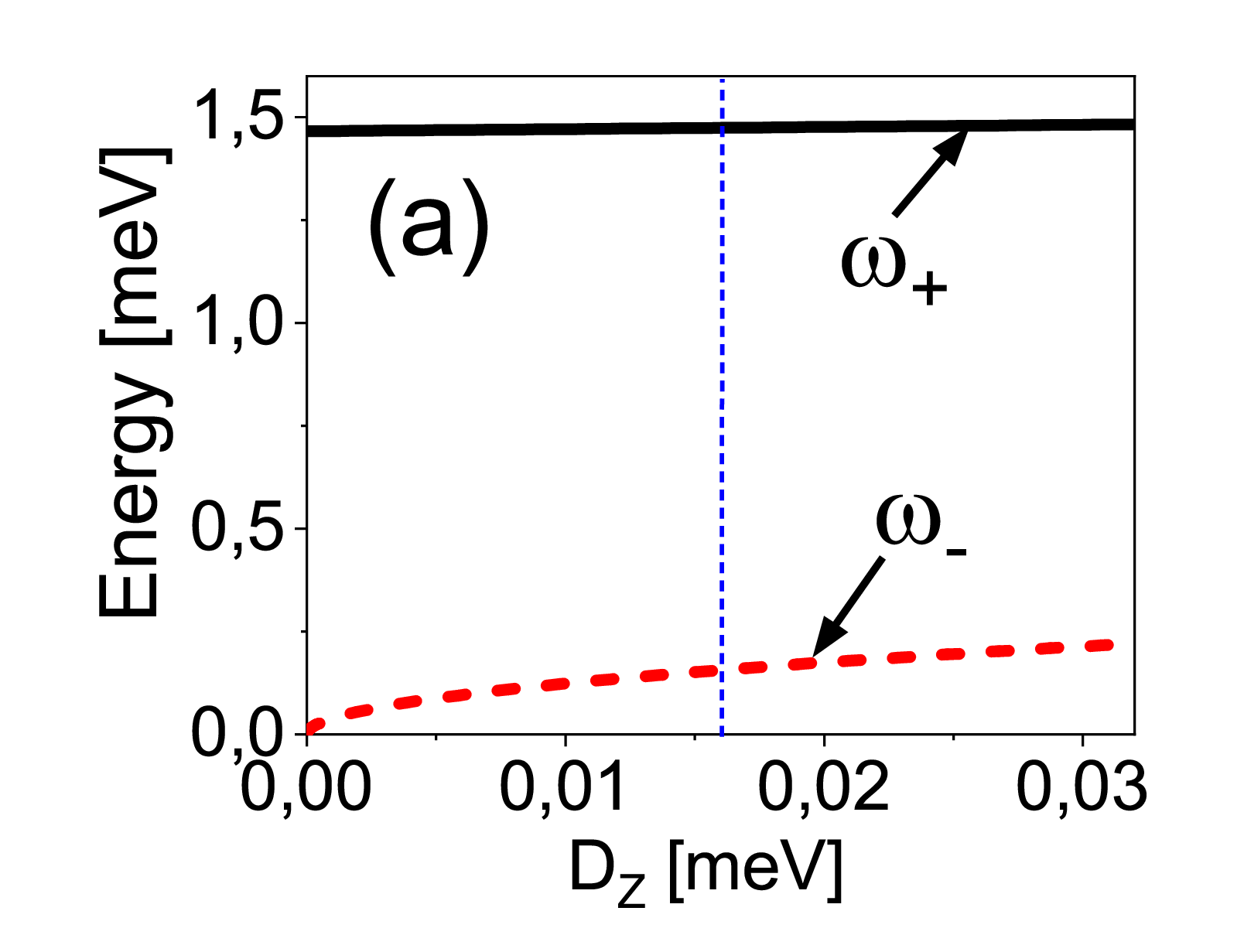}\\
    \includegraphics[width=0.95\columnwidth]{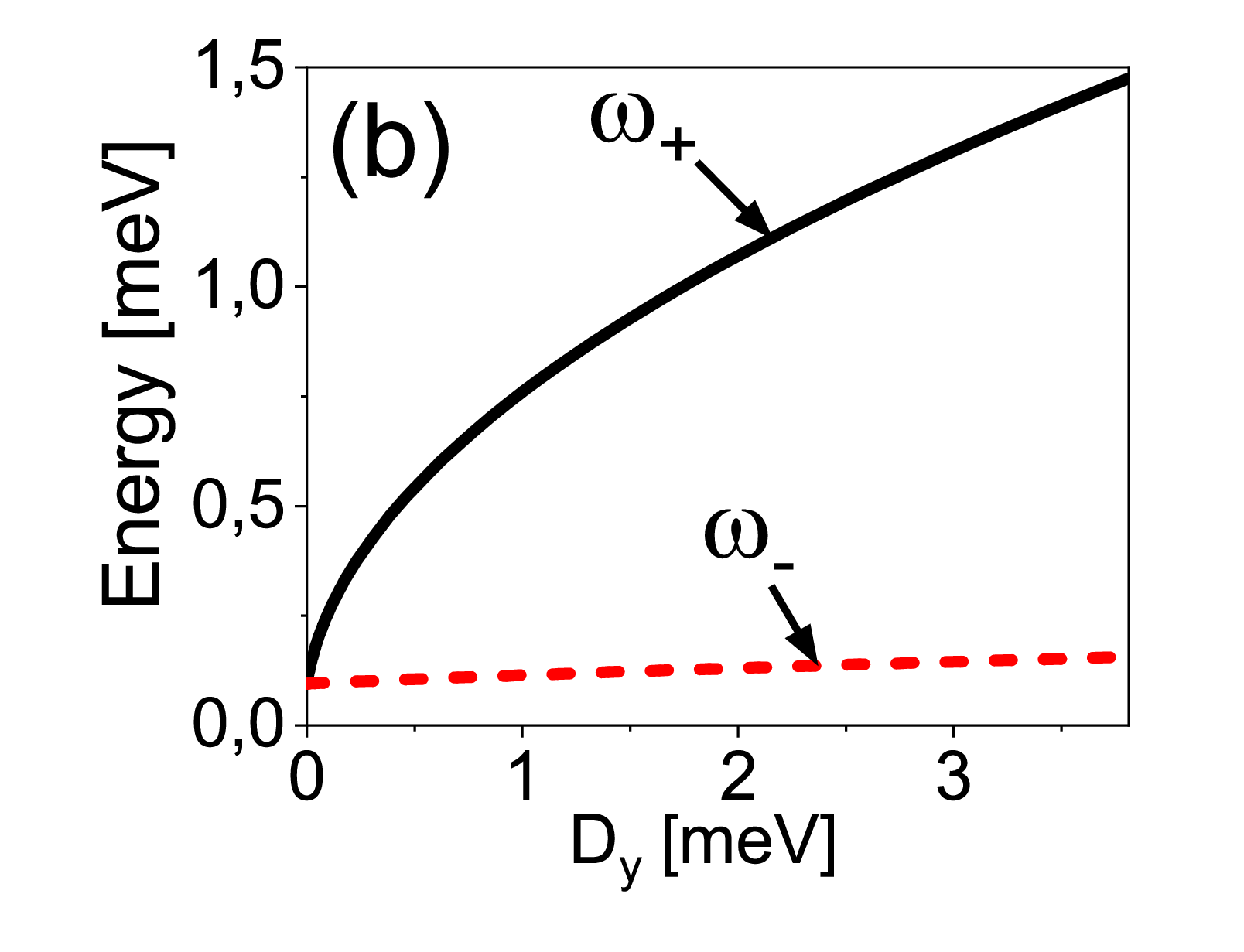}
\caption{Energy of the two modes in the $\Gamma$ point as a function of (a) the in-plane  easy-axis anisotropy constant $D_z$ (the blue dashed line indicates $D_z$ used for VTe$_2$ system), and (b) as a function of the easy-plane magnetic anisotropy constant $D_y$.}
    \label{Fig:4}
\end{figure}

Explicit dispersion relations along  the $\Gamma\to K$ and $\Gamma\to M\to K$ paths in the Brillouin zone are shown in Fig.3a. Thoough there are two modes in the bilayers under considerations, separation of these two  modes is not resolved in Fig.3a. Therefore, in Fig.3b we zoomed in a small area near the $\Gamma$ point. Now, splitting of the modes is clearly seen.  Moreover, this figure also shows that there is a gap in the spectrum at the $\Gamma$ point, i.e. the spin wave energy does not vanish at $k=0$. This gap is  a consequence of the in-plane easy-axis anisotropy, $D_z$, as shown in Fig.4a, where the two modes at the $\Gamma$ point are  plotted as a function of $D_z$. When $D_z=0$, energy of the lowest mode vanishes, while the gap between the two modes survives. This gap, in turn, is determined mainly by the easy plane anisotropy constant, as shown in Fig.4b.

As we have already discussed above, external magnetic field leads to spin reorientation in the two monolayers. The transition to spin-flop phase  appears at $h_{\rm sf}$  and then at the saturation field $h_s$ the transition to fully collinear (ferromagnetic) phase takes place. The  threshold fields $h_s$ and $h_{\rm sf}$ are determined by the anisotropy constants and interlayer exchange parameter, as described in section 2. Variation of the the fields  $h_s$ and $h_{\rm sf}$ with the easy-axis anisotropy constant $D_z$ is shown in Fig.5.  To present a complete physical picture, we plot there $h_s$ and $h_{\rm sf}$ for $D_z$ significantly exceeding the  value of $D_z$ in 2H-VTe$_2$, while the other parameters correspond to  2H-VTe$_2$. This figure shows that when $D_z=0$, the transition to the spin-flop phase appears already at infinitesimally small magnetic field. This figure also shows that in a general case the saturation field decreases with increasing $D_z$ and for a specific value of $D_z$ (denoted as  $D_z^M$) the saturation field and the transition field to the spin-flop phase become equal, and  the system changes from antiferromagnetic to ferromagnetic without the intermediate spin-flop phase (so-called metamagnetic transition). However, this does not happen in 2H-VTe$_2$ due to a small value of $D_z$ in this material, indicated by the vertical dashed line in Fig.5. Accordingly, transition to the spin-flop phase takes place at a small magnetic field, and the range of magnetic field where the antiparallel configuration is stable is very narrow. In turn, the range of magnetic field with stable spin-flop phase is relatively large. This is shown explicitly in Fig.6, where the dispersion curves are plotted as a function of external magnetic field.  

\begin{figure}[hbt!]
\centering
    \includegraphics[width=0.95\columnwidth]{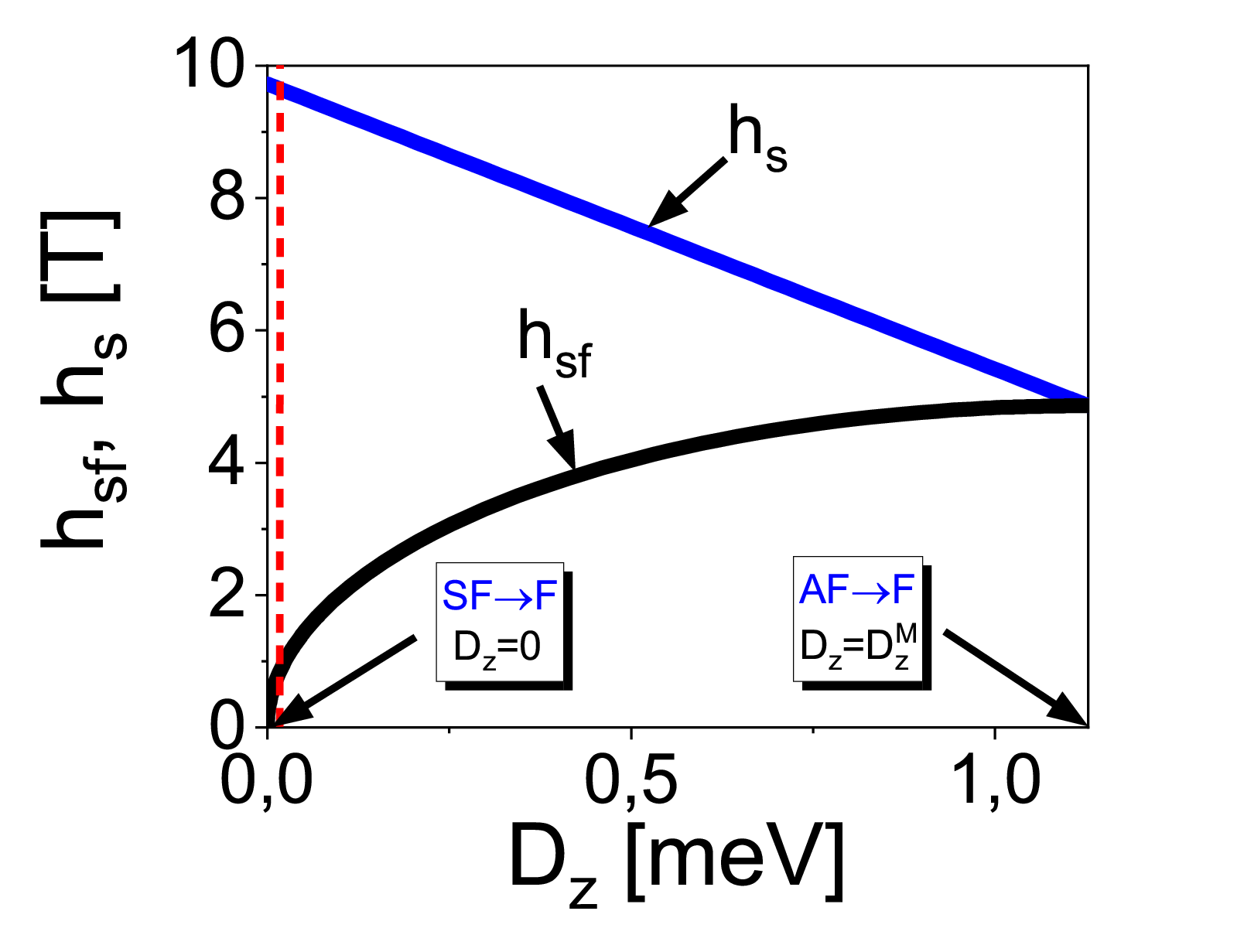}
\caption{The spin-flop field $h_{\rm sf}$ and saturation field $h_s$, plotted  as a function of in-plane easy-axis anistropy constant $D_z$.  The range of $D_z$ in this figure covers the full range where the spin-flop phase may appear, i.e. from  $D_z=0$ to $D_z$ where a metamagnetic transition occurs.  The vertical dashed line corresponds to $D_z$ in VTe$_2$, which indicates that the range of magnetic field where the antiparallel state exist is very narrow, while the range of field with stable spin-flop phase is relatively broad. Note, the field units are here converted from energy units to  Tesla.}  
\end{figure}

\begin{figure}[hbt!]
\centering
    \includegraphics[width=0.95\columnwidth]{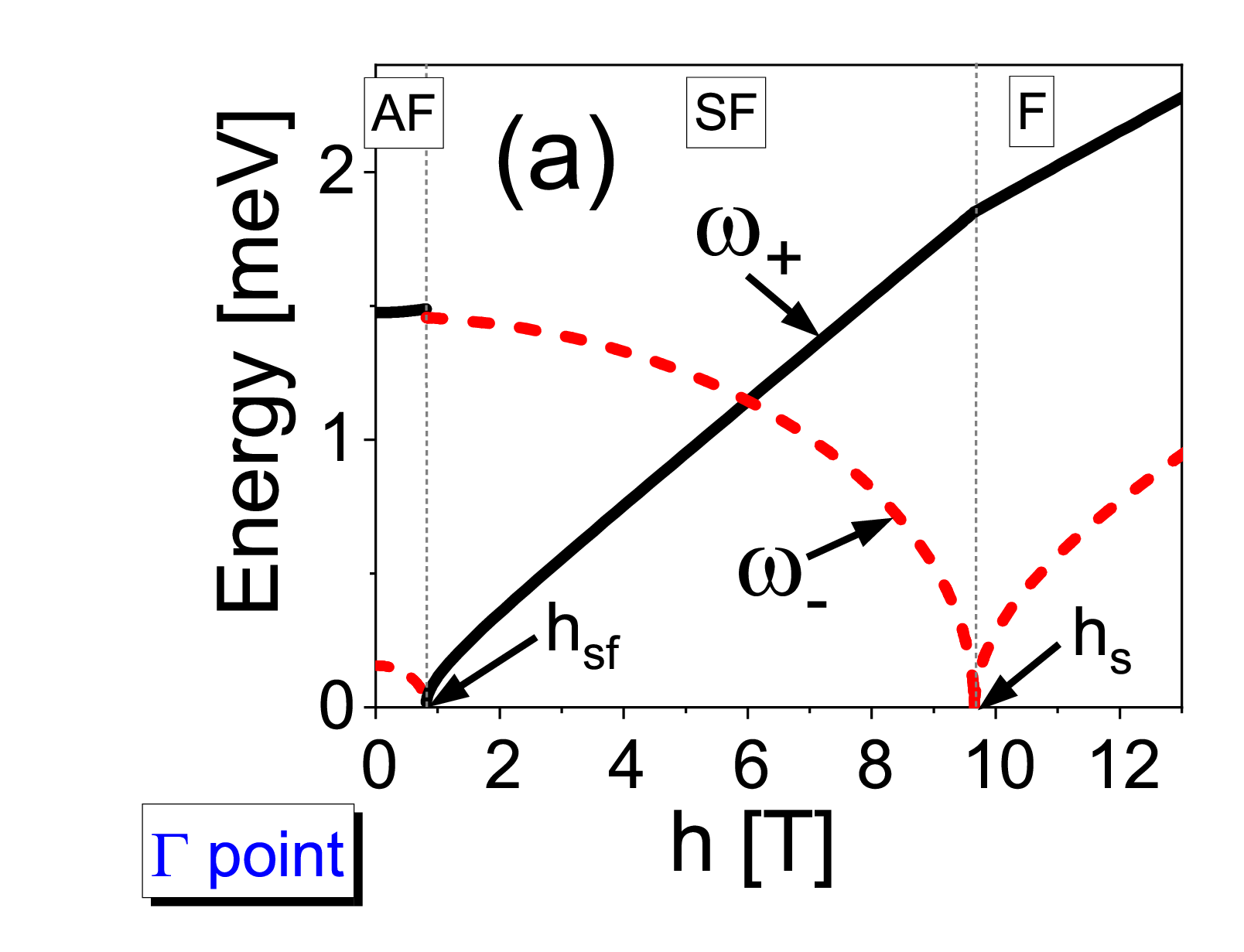}\\
    \includegraphics[width=0.95\columnwidth]{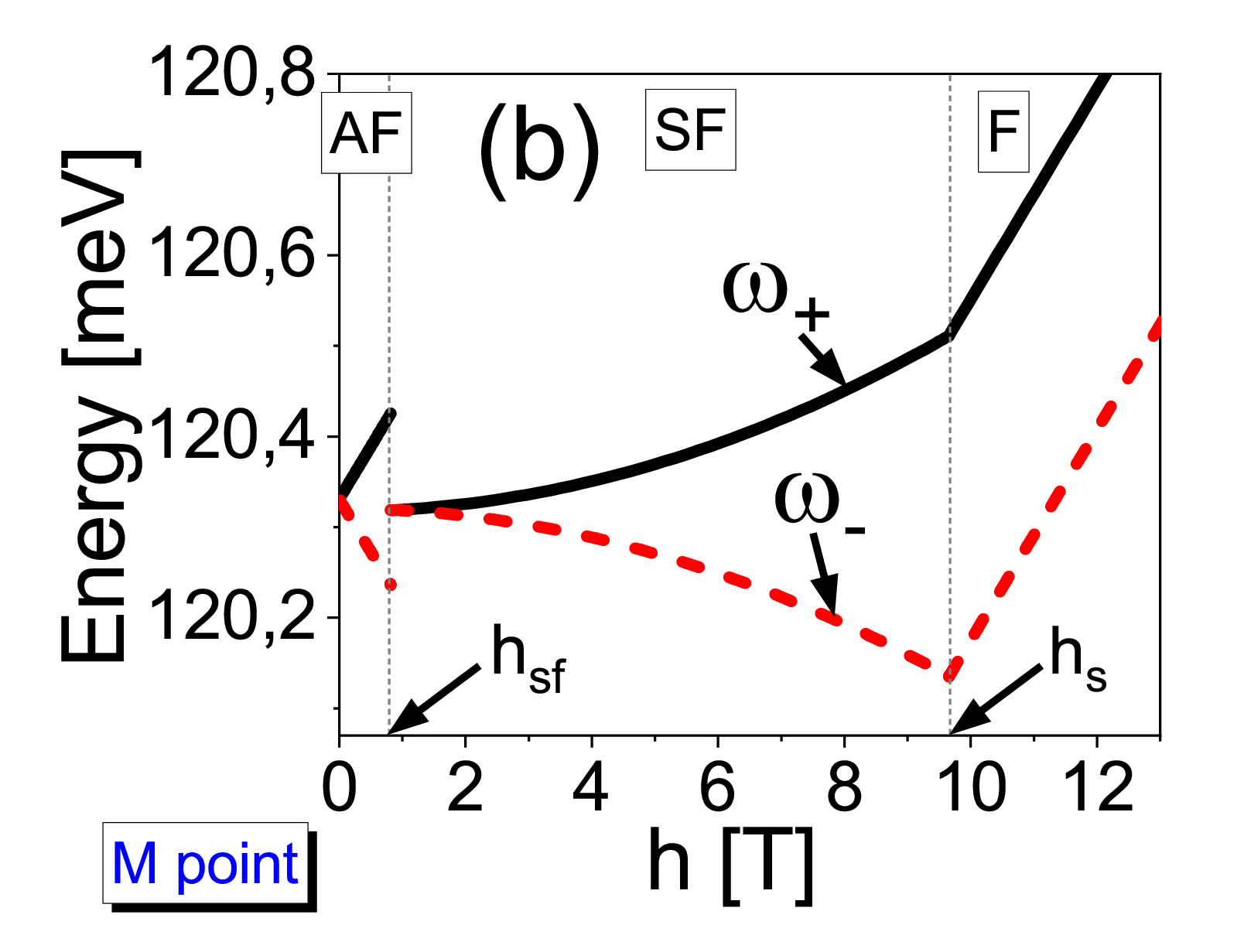}\\
    \includegraphics[width=0.95\columnwidth]{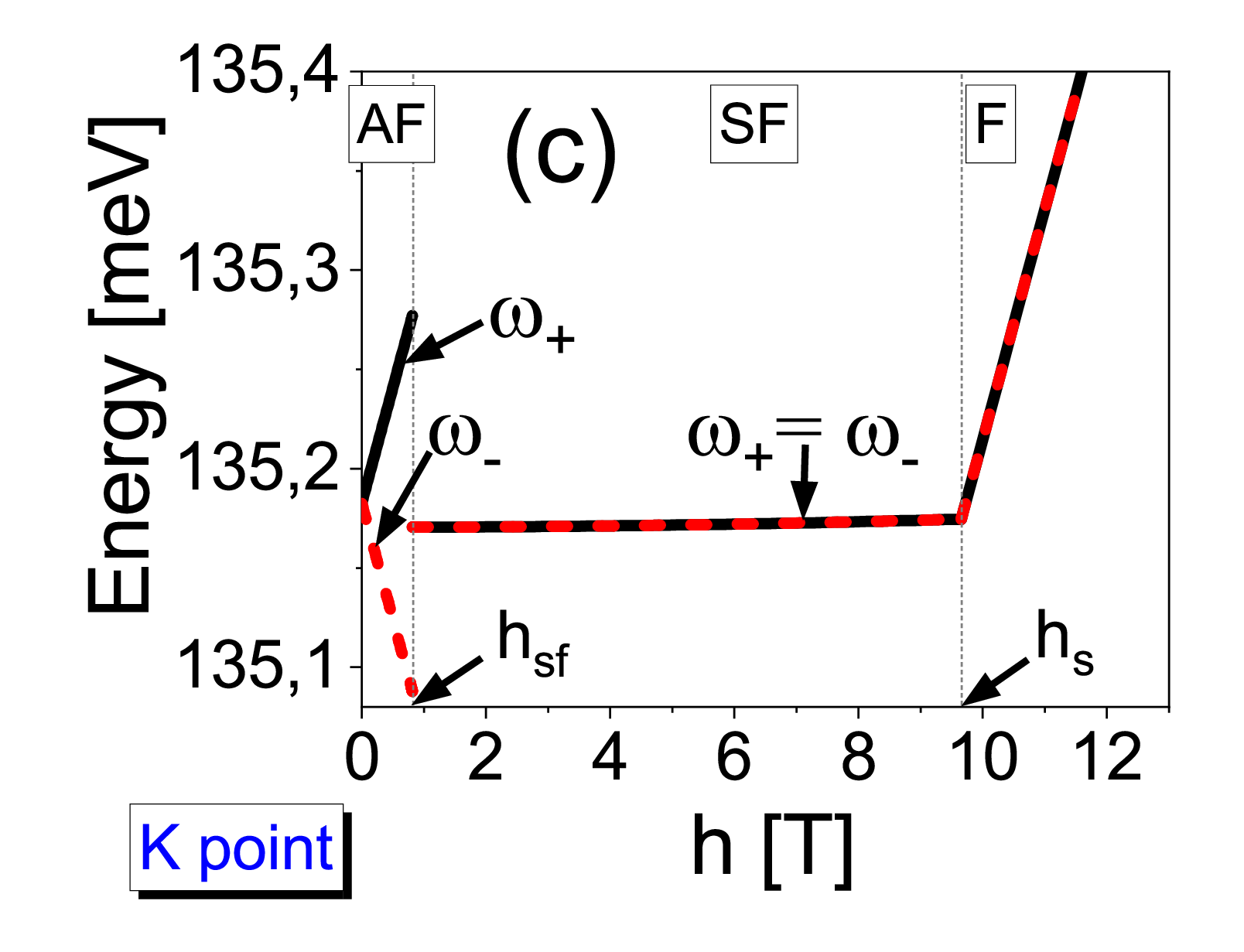}\\
\caption{Magnetic field ($h$) dependence of the spin wave spectra in the $\Gamma$ (a), M (b), and K (c) points of the Brillouin zone. With increasing $h$, system goes from antiferromagnetic (AF) state to the spin-flop (SF) state at $h=h_{\rm sf}$, and then  from the spin-flop phase to the ferromagnetic (FM) state at $h=h_s$. Note, the field units are here converted from energy units to  Tesla.}
\end{figure}

Variation of the spin wave energy at the $\Gamma$ point with increasing external magnetic field is shown in Fig.6a, while the same for the M and K points is shown in Fig.6b and Fig.6c, respectively. Note, that the two modes (black and red curves), behave differently in the three regions of magnetic field. In the antiferromagnetic phase, energy of one of the modes (that of higher energy) slightly increases with the field, wheres  the second mode (of lower energy) becomes softened and reaches a minimum at the transition point to the spin-flop phase. This minimum is nonzero due to a finite in-plane easy-axis anisotropy. Slightly above the spin-flop field $h_{\rm sf}$, energy of the lower mode (the soft one) increases with the field $h$, while energy of the upper mode decreases with increasing $h$. This tendency is kept till the saturation field $h_s$, where the latter modes becomes soft and its energy  vanishes at $h_s$, while the energy of the former mode still increases with $h$ for $h>h_s$, however a clear kink appears at the transition between the spin-flop and ferromagnetic phases.   Similar behavior of the spin waves in the K and M points of the Brillouin zone is shown in Fig.6b and Fig.6c.  In the antiferromagnetic phase one mode increase linearly with magnetic field, while the other decrease linearly with $h$. This is valid for both, K and M points. Interestingly, the two modes in the K point are degenerate in the spin-flop phase as well as in the ferromagnetic phase.

We note that some features of the spinwave spectra are similar to those found in spinwave spectra of  artificially layered structures with antiferromagnetic interlayer coupling. Magnetic and electronic transport properties of such structures were studied extensively in 80s and 90s, and the efforts resulted in the discovery of giant magnetoresistance effect~\cite{Baibich88,Binash89}, which follows from spin dependent scattering processes~\cite{Camley89}. The magnetic van der Waals materials are also layered magnetic structures, but with atomically thin magnetic layers and no nonmagnetic spacing layers. Therefore,  spinwave spectra in both types of materials, i.e., artificial and van der Waals layerd magnetic structures, have some similarities, and this similarity is especially evident in microscopic calculations of spin waves in artificially layered systems, like in the paper by Nortemann {\it et al}~\cite{Nortemann}.
In that paper, the authors considered a stack of magnetic layers that were coupled antiferromagnetically, and in each magnetic layer there was a cubic spin lattice. Including exchange interactions, dipolar  coupling, and Zeeman energy in an external magnetic field, the authors evaluated the spinwave spectra  using the linearized equation of motion technique.
In our paper we consider the bilayer of Vanadium based dichalcogenides with antiferromagnetic coupling of the two monolayers, and  spin lattice in individual monolayers is hexagonal as shown in Fig.1. We also include  two anisotropies, easy-plane and easy-axis ones, and calculate spinwaves  within a fully quantum mechanical approach based on the Hollstein-Primakoff transformation followed by the Bogoliubov transformation.
Despite some differences in the systems and techniques, certain features of the spinwave spectra are qualitatively similar, including softening of one spinwave mode at the  transition form the antiferromagnetic to  the spin-flop phase and at the transition from spin-flop to parallel state.
Finally, we mention that various models and approximations were used to calculate spinwaves in artificially layerd structures, including microscopic and  macroscopic descriptions, effective medium theory, and others \cite{Stamps,Barnas}.

\section{Summary}

In this paper we have analyzed spin wave modes in a class of van der Waals magnetic materials, that includes transition-metal dichalcogenides. The description is limited to bilayers with easy plane anisotropy and ferromagnetic intralayer exchange coupling, i.e.  individual monolayers are ferromagnetically ordered in the layer plane. In turn, the two layers are coupled either ferromagnetically or antiferromagnetically. To find the spin wave energy we used the Hollstein-Primakoff-Bogolubov diagonalization scheme.

The bilayers support two magnon modes, which are split in general, though the splitting is rather small due to small interlayer exchange coupling. In the absence of external magnetic field and in-plane easy-axis anisotropy, energy of one of  the modes vanishes in the $\Gamma$ point of the Brillouin zone. This mode is the well known Goldstone mode. External field or in-plane anisotropy create a gap at the $\Gamma$ point.

Van der Waals materials are of current interest from the point of view of possible applications.  But of particular interest are bilayers of van der Waals magnetic materials, that can be considered  as natural atomically-thin spin valves.

\acknowledgements
{This work has been supported by the Norwegian Financial Mechanism 2014- 2021 under the Polish – Norwegian Research Project NCN GRIEG “2Dtronics” no. 2019/34/H/ST3/00515.}

\appendix
\section{Spin phases}

As already mentioned in the main text, for the model Hamiltonian (1) and (2)  one may expect in general three  stable spin configurations of the bilayer system in an external magnetic field applied along the in-plane easy axis, i.e. (i) antiferromagnetic state at low fields with the spins of the two monolayers oriented along $+z$ and $-z$ axis for the bottom and top layers, respectively, (ii) spin-flop phase in a specific range of magnetic field, with the spins of the two monolayers lying in the atomic planes at an angle $\chi$ to the $z$ axis, (iii)  and the ferromagnetic  phase with the spins of both layers along the $z$ axis.
In order to determine these phases in a specific magnetic field (and also bearing in mind the magnon description)  it is convenient to use coordinate systems with the local $z'$ axes along the corresponding spin orientations.
To do this one has to combine rotation of the  spins from the global frame around the y-axis by the canting angle $\chi$ and around the z-axis by the angle $\theta_\alpha$, where  $\alpha=T$ and $\alpha=B$ labels the top and bottom layers, respectively,
\begin{equation}
    \textbf{S}_{\textbf{r},\alpha}={\mathbf{\hat{R}}_z}(\theta_\alpha){\mathbf{\hat{R}}_y}(\chi)\textbf{S}^{'}_{\textbf{r},\alpha},
\end{equation}
where the rotation matrix  reads
\begin{equation}
{\mathbf{\hat{R}}_z}(\theta_\alpha){\mathbf{\hat{R}}_y}(\chi)=
    \left( \begin{array}{ccc}
\cos{\theta_\alpha}\cos{\chi} & -\sin{\theta_\alpha} & \cos{\theta_\alpha}\sin{\chi}\\
\sin{\theta_\alpha}\cos{\chi} & \cos{\theta_\alpha} & \sin{\theta_\alpha}\sin{\chi}\\
-\sin{\chi} & 0 & \cos{\chi}\\
\end{array} \right),
\end{equation}
with $\theta_T=\pi$ for the top layer, $\theta_B=0$ for the bottom layer, and $\chi$ being the polar angle between the spin (aligned along the $z'$-axis of the local  coordinate system) and the $z$-axis of the global   frame. Thus,
\begin{eqnarray}
    S_{\textbf{r},\alpha}^x={\mp}S_{\textbf{r},\alpha}^{'x}\sin{\chi}{\mp}S_{\textbf{r},\alpha}^{'z}\cos{\chi}, \\
    S_{\textbf{r},\alpha}^y={\mp}S_{\textbf{r},\alpha}^{'y}, \\
    S_{\textbf{r},\alpha}^z=-S_{\textbf{r},\alpha}^{'x}\cos{\chi}+S_{\textbf{r},\alpha}^{'z}\sin{\chi},
\end{eqnarray}
where the sign $-(+)$ corresponds to the layers $\alpha=T$ ($\alpha=B$), respectively.

The $h$- dependent regimes of the spin configurations of the bilayer can be found  from the classical energy. In the spin-flop phase, $h_{\rm sf}\leq{h}\leq{h_s}$
\begin{eqnarray}
    E_{\rm sf}/NS & = & -6|J_1|S+{\xi}J_2S\cos2\chi-\frac{1}{2}D_zS\cos^2\chi
    \nonumber\\
    & & -h\cos{\chi},
\end{eqnarray}
where $N$ is the total number of sites, and $\xi$ denotes the structure factor: $\xi =3$ and $\xi =1$ for the H and T phase, respectively.
Hence,  minimizing the classical energy, ${\partial}E_{\rm sf}/{\partial\chi}=0$, yields the condition for the canting angle $\chi$,
   $ \cos{\chi}={h}/{h_s}$,
where $h_s$ is the threshold magnetic field (the saturation field) at which the transition between spin-flop   and ferromagnetic ($\chi=0$) phases occurs,
\begin{equation}
h_s=S(4{\xi}J_2-D_z).
\end{equation}
The threshold $h_{\rm sf}$ for transition from the antiferromagnetic phase to the  spin-flop one can be derived from the condition  $E_{sf}=E_{\rm af}$, where $E_{\rm af}$ denotes the classical energy for the collinear antiferromagnetic phase,
\begin{equation}
    E_{\rm af}/NS=-6|J_1|S-{\xi}J_2S-\frac{1}{2}D_zS.
\end{equation}
Thus, from Eqs. (A6) to (A8) one finds
\begin{equation}
    h_{\rm sf}=\sqrt{SD_zh_s}=S\sqrt{D_z(4{\xi}J_2-D_z)}.
\end{equation}
One can see that $h_{\rm sf}=0$ if $D_z=0$. Thus, even for nonvanishing magnetic fields, $0<h<h_{sf}$, the collinear antiferromagnetic configuration  may exist as it is stabilized by the  in-plane magnetic anisotropy.

\section{Structural properties}

For the considered VSe$_2$ system, each layer has hexagonal lattice with the primitive lattice vectors
\begin{equation}
    \textbf{a}_{1,2}=a\bigg({\pm}\frac{\sqrt{3}}{2}\textbf{\^{x}}+\frac{1}{2}\textbf{\^{z}}\bigg),\> \textbf{a}_3=0,
\end{equation}
where $a$ is the in-plane lattice constant (distance between Vanadium atoms). Here,  each  V atom has six intralayer nearest neigbours determined by the $\bm{\delta}$ vectors:
\begin{eqnarray}
    &&\bm{\delta}_{1,2}={\pm}a\textbf{\^{z}},
    \nonumber\\
    & &
    \bm{\delta}_{3,4}=a\bigg({\pm}\frac{\sqrt{3}}{2}\textbf{\^{x}}+\frac{1}{2}\textbf{\^{z}}\bigg),
    \nonumber\\
    & &
     \bm{\delta}_{5,6}=a\bigg({\pm}\frac{\sqrt{3}}{2}\textbf{\^{x}}-\frac{1}{2}\textbf{\^{z}}\bigg).
\end{eqnarray}
Moreover, for the T-stacked system, each V atom has one NN in the adjacent layer, while for the H-stacked system there are three NNs in the adjacent layer with $\bm{\delta}$ given by
\begin{equation}
    \bm{\delta}_{1,3}=a\bigg(-\frac{1}{2\sqrt{3}}\textbf{\^{x}}{\pm}\frac{1}{2}\textbf{\^{z}}\bigg),{\>}\bm{\delta}_2=\frac{a}{\sqrt{3}}\textbf{\^{x}}.
\end{equation}

\section{Hollstein-Primakoff transformations in SF phase}

Using the Holstein-Primakoff transformation, which for the SF configuration reads
\begin{equation}
    S_{\textbf{r},\alpha}^x=\sqrt{\frac{S}{2}}(a_{\textbf{r},\alpha}^++a_{\textbf{r},\alpha}),
\end{equation}
\begin{equation}
    S_{\textbf{r},\alpha}^y=i\sqrt{\frac{S}{2}}(a_{\textbf{r},\alpha}^+-a_{\textbf{r},\alpha}),
\end{equation}
\begin{equation}
    S_{\textbf{r},\alpha}^z=S-a_{\textbf{r},\alpha}^+a_{\textbf{r},\alpha},
\end{equation}
we arrive at the following form of the Hamiltonian written for the bosonic operators:
\begin{eqnarray}
    H & = &
    J_1S\sum_{\textbf{r},\bm{\delta},\alpha}\big(a_{\textbf{r},\alpha}^+a_{\textbf{r}+\delta,\alpha}+a_{\textbf{r}+\delta,\alpha}^+a_{\textbf{r},\alpha}
    \nonumber\\
    & &   -a_{\textbf{r},\alpha}^+a_{\textbf{r},\alpha}-a_{\textbf{r}+\delta,\alpha}^+a_{\textbf{r}+\delta,\alpha}\big)
    \nonumber\\
    & & +2J_2S\sum_{\textbf{r},\bm{\delta}}\Big[-\cos2\chi\big(a_{\textbf{r},T}^+a_{\textbf{r},T}+a_{\textbf{r}+\delta,B}^+a_{\textbf{r}+\delta,B}\big)
\nonumber\\
    & &+\sin^2\chi\big(a_{\textbf{r},T}^+a_{\textbf{r}+\delta,B}^++a_{\textbf{r},T}a_{\textbf{r}+\delta,B}\big)
\nonumber\\
    & &-\cos^2\chi\big(a_{\textbf{r},T}^+a_{\textbf{r}+\delta,B}+a_{\textbf{r},T}a_{\textbf{r}+\delta,B}^+\big)\Big]
    \nonumber\\
    & & +\frac{D_yS}{4}\sum_{\textbf{r},\bm{\alpha}}\Big(2a_{\textbf{r},\alpha}^+a_{\textbf{r},\alpha}-a_{\textbf{r},\alpha}^+a_{\textbf{r},\alpha}^+-a_{\textbf{r},\alpha}a_{\textbf{r},\alpha}\Big)
     \nonumber\\
&&+\frac{D_zS}{2}\sum_{\textbf{r},\bm{\alpha}}\Big[\big(3\cos^2\chi-1\big)a_{\textbf{r},\alpha}^+a_{\textbf{r},\alpha}
    \nonumber\\
    & &
    -\frac{1}{2}\sin^2\chi\big(a_{\textbf{r},\alpha}^+a_{\textbf{r},\alpha}^++a_{\textbf{r},\alpha}a_{\textbf{r},\alpha}\big)\Big]
    \nonumber\\
    & &
    +h\cos\chi\sum_{\textbf{r},\alpha}a_{\textbf{r},\alpha}^+a_{\textbf{r},\alpha}.
\end{eqnarray}
The Fourier transformation described by Eq. (9) together with Eqs. (B1)-(B3) yields
\begin{eqnarray}
    H & = & \sum_{\textbf{k}}\bigg{\{}2J_1S\sum_{\alpha}(\gamma_\textbf{k}-6)a_{\textbf{k},\alpha}^+a_{\textbf{k},\alpha}
    \nonumber\\
    & & +2J_2S\Big[-\xi\cos2{\chi}\sum_{\alpha}a_{\textbf{k},\alpha}^+a_{\textbf{k},\alpha}
\nonumber\\
    & &+\sin^2\chi\big(\eta_{\textbf{k}}a_{\textbf{-k},T}^+a_{\textbf{k},B}^++\eta_{\textbf{k}}^*a_{\textbf{-k},T}a_{\textbf{k},B}\big)
\nonumber\\
    & &-\cos^2\chi\big(\eta_{\textbf{k}}^*a_{\textbf{k},T}^+a_{\textbf{k},B}+\eta_{\textbf{k}}a_{\textbf{k},T}a_{\textbf{k},B}^+\big)\Big]
    \nonumber\\
    & &+\frac{S}{2}\sum_{\bm{\alpha}}\bigg{\{}\Big[D_y+D_z(3\cos^2\chi-1)\bigg]a_{\textbf{k},\alpha}^+a_{\textbf{k},\alpha}
\nonumber\\
    & &
    -\frac{1}{2}\big(D_y+D_z\sin^2\chi\big)\big(a_{\textbf{k},\alpha}^+a_{\textbf{-k},\alpha}^++a_{\textbf{k},\alpha}a_{\textbf{-k},\alpha}\big)\bigg{\}}
    \nonumber\\
    & &+h\cos\chi\sum_{\alpha}a_{\textbf{k},\alpha}^+a_{\textbf{k},\alpha}\bigg{\}},
\end{eqnarray}
which within the Bogolubov transformation approach leads to the final form of the Hamiltonian $H=H_{\textbf{k}}+H_{\textbf{-k}}$, where $H_{\textbf{k}}$ is given by Eq.(33) in the main text.

\section{Ferromagnetic interlayer coupling }

If the TMD 2H-VX$_2$ bilayer has FM ground state (as e.g. 2H-VS$_2$), then Eq. (3) describes ferromagnetic interlayer coupling with exchange coupling parameter $J_2<0$.
In such a case the  Holstein-Primakoff and Fourier transformations for the top and bottom layers
lead to the full Hamiltonian in the form
\begin{eqnarray}
    H & = & \sum_{\textbf{k}}\bigg{\{}2J_1S\sum_{\alpha}(\gamma_\textbf{k}-6)a_{\textbf{k},\alpha}^+a_{\textbf{k},\alpha}
    \nonumber\\
    & & +2J_2S\Big(-{\xi}\sum_{\alpha}a_{\textbf{k},\alpha}^+a_{\textbf{k},\alpha}+\eta_{\textbf{k}}a_{\textbf{k},T}a_{\textbf{k},B}^++\eta_{\textbf{k}}^*a_{\textbf{k},T}^+a_{\textbf{k},B}\Big)
    \nonumber\\
    & &+S\sum_{\alpha}\Big[-\frac{D_y}{4}\big(a_{\textbf{k},\alpha}^+a_{-\textbf{k},\alpha}^++a_{\textbf{k},\alpha}a_{-\textbf{k},\alpha}-2a_{\textbf{k},\alpha}^+a_{\textbf{k},\alpha}\big)
    \nonumber\\
    & &
    +D_za_{\textbf{k},\alpha}^+a_{\textbf{k},\alpha}\Big]+h\sum_{\alpha}a_{\textbf{k},\alpha}^+a_{\textbf{k},\alpha}\bigg{\}}.
\end{eqnarray}
This Hamiltonian can be written as
\begin{equation}
    H=H_{\textbf{k}}+H_{\textbf{-k}},
\end{equation}
where
\begin{eqnarray}
    H_{\textbf{k}} & = & \sum_{\textbf{k}}\bigg[\sum_{\alpha}\bigg(\frac{A_{\textbf{k}}}{2}\bigg)a_{\textbf{k},\alpha}^+a_{\textbf{k},\alpha}
    +B_{\textbf{k}}a_{\textbf{k},T}^+a_{\textbf{k},B}
    \nonumber\\
    & &
    + C\sum_{\alpha}a_{\textbf{k},\alpha}a_{\textbf{-k},\alpha}\bigg]+H.c.,
\end{eqnarray}
with
\begin{eqnarray}
    &&A_{\textbf{k}}=S\Big[2J_1\big(\gamma_\textbf{k}-6\big)-2{\xi}J_2+\frac{D_y}{2}+D_z\Big]+ h,
    \nonumber\\
    & &
    B_{\textbf{k}}=2{\eta}^*_{\textbf{k}}J_2S,
    \nonumber\\
    & &
     C=-\frac{D_yS}{4}.
\end{eqnarray}
 The eigenvalue problem evaluated by means of the Bogolubov diagonalization scheme
leads finally to the dispersion relation
given by the formula
\begin{eqnarray}
    \omega_{\bm{k},\mu} =
    \Big[\big(A_{\bm{k}}\pm|B_{\bm{k}}|\big)^2-4C^2\Big]^{\frac{1}{2}},
\end{eqnarray}
where $A_{\bm{k}}$ and $|B_{\bm{k}}|$ are given by Eq. (D4) (note that here $|B_{\bm{k}}|\equiv|\tilde{B}_{\bm{k}}|$), and where $\pm$ corresponds to mode $\mu=+, -$. At the zone center, $\bm{k}=0$, one gets
\begin{equation}
    \omega_{\bm{k}=0,+} =S\bigg[\Big(12|J_2|+\frac{D_y}{2}+D_z\Big)^2-\Big(\frac{D_y}{2}\Big)^2\bigg]^\frac{1}{2}
\end{equation}
\begin{equation}
    \omega_{\bm{k}=0,-} =S\bigg[\Big(\frac{D_y}{2}+D_z\Big)^2-\Big(\frac{D_y}{2}\Big)^2\bigg]^\frac{1}{2},
\end{equation}
so that in the absence of the Zeeman field ($h=0$) as well as in the absence of the in-plane anisotropy field ($D_z=0$), one finds $\omega_{\bm{k}=0,-}=0$. In such a case, for $D_y>0$ a gapless, linearly vanishing $\sim|\bm{k}|$ Goldstone
mode occurs at the zone center, while for $D_y=0$  the mode $\omega_{\bm{k},-}$ vanishes non-linearly in the vicinity of $\bm{k}=0$.
\\

\end{document}